\documentclass{article}

\usepackage[margin=1in]{geometry} 
\usepackage{natbib}               
\usepackage{graphicx}             
\usepackage{amsmath, amssymb}     
\usepackage{url}                  
\usepackage{authblk}              

\usepackage[utf8]{inputenc} 
\usepackage[T1]{fontenc}    
\usepackage{url}            
\usepackage{booktabs}       
\usepackage{amsfonts}
\usepackage{wrapfig}       
\usepackage{nicefrac}       
\usepackage{microtype}      
\usepackage{pifont}
\usepackage{booktabs}
\usepackage[table]{xcolor} 
\usepackage{mathrsfs}
\usepackage{graphicx}
\usepackage{enumitem}
\usepackage{amsmath}
\usepackage{graphicx}    
\usepackage{subcaption}  
\usepackage[most]{tcolorbox}
\usepackage{tabularx}
\usepackage{booktabs}
\usepackage{multirow}
\usepackage{algorithm}
\usepackage{algpseudocode}
\usepackage{amsmath}
\usepackage{hyperref}    

\title{MasHost Builds It All: Autonomous Multi-Agent System Directed by Reinforcement Learning}

\author{
	Kuo Yang$^{1,2}$, Xingjie Yang $^{1}$, Linhui Yu$^{1}$, Qing Xu$^{1}$, Yan Fang$^{1,3}$, Xu Wang$^{1,2}$, \\
	Zhengyang Zhou$^{1,2}$ \textsuperscript{$\ast$}, Yang Wang$^{1,2}$ \thanks{\noindent Zhengyang Zhou and Yang Wang are corresponding authors.}\\
	$^1$University of Science and Technology of China (USTC), Hefei, China\\
	$^2$Suzhou Institute for Advanced Research, USTC, Suzhou, China\\
	$^3$Hong Kong University of Science and Technology (Guangzhou), Guangzhou, China\\
	yangkuo@mail.ustc.edu.cn
}

\date{} 
\begin{document}

\maketitle

\begin{abstract}
Large Language Model (LLM)-driven Multi-agent systems (Mas) have recently emerged as a powerful paradigm for tackling complex real-world tasks. However, existing {Mas} construction methods typically rely on manually crafted  interaction mechanisms or heuristic rules, introducing human biases and constraining the autonomous ability. Even with recent advances in adaptive Mas construction, existing systems largely remain within the paradigm of semi-autonomous patterns. In this work, we propose \texttt{MasHost}, a Reinforcement Learning (RL)-based framework for autonomous and query-adaptive Mas design.  By formulating Mas construction as a graph search problem, our proposed \texttt{MasHost} jointly samples agent roles and their interactions through a unified probabilistic sampling mechanism. Beyond the accuracy and efficiency objectives pursued in prior works, we introduce component rationality as an additional and novel design principle in {Mas}. To achieve this multi-objective optimization, we propose Hierarchical Relative Policy Optimization (HRPO), a novel RL strategy that collaboratively integrates group-relative advantages and action-wise rewards. To our knowledge, our proposed \texttt{MasHost} is the first RL-driven framework for autonomous Mas graph construction. Extensive experiments on six benchmarks demonstrate that \texttt{MasHost} consistently outperforms most competitive baselines, validating its effectiveness, efficiency, and structure rationality. \footnote{The code will be released upon acceptance of the paper.} 
\end{abstract}

\section{Introduction}
\label{Introduction}

In recent years, the advent of large language models (LLMs) has fundamentally reshaped research paradigms across various fields \cite{achiam2023gpt, shao2024deepseekmath, openai2024gpt4omini}.
LLM-driven Multi-agent system ({Mas}) demonstrate remarkable potential in addressing complex real-world tasks, emerging as a prominent research frontier in artificial intelligence \cite{ zhuge2024gptswarm, zhang2024g, zhang2024aflow, wang2023unleashing, yang2023auto, hong2023metagpt, liu2023dynamic, wu2023autogen, ye2025mas, zhang2025multi, hu2024automated, chen2023autoagents}.  {Mas} seeks to address tasks that surpass the capabilities of a single agent through coordinated interactions among multiple agents \cite{luo2025large, li2024personal, guo2024large}. Therefore, designing the interaction mechanism among agents is critical to ensuring the effectiveness of {Mas}.  Many studies rely on manual drafting and heuristic-based approaches for constructing interaction mechanisms \cite{wei2022chain, wang2023unleashing, du2023improving, wang2022self}. However, these strategies often yield suboptimal performance due to the introduction of human biases.


This limitation has prompted recent efforts toward the development of autonomous {Mas}. These works model {Mas} as a directed graph to achieve policy-driven Mas construction, facilitating more adaptive and flexible connections among agents \cite{zhuge2024gptswarm, zhang2024g, zhang2024aflow, hong2023metagpt, hu2024automated, chen2023autoagents, yue2025masrouter, zhang2025multi}. 
Despite these advances, full autonomous {Mas} remains elusive.  \ding{182} \textbf{Candidate Pool Sampling} strategy is followed by many existing approaches  \cite{zhang2025multi, yue2025masrouter, chen2023autoagents}, where {Mas} are constructed by sampling or composing from a predefined structure pool. This candidate pool inevitably introduces human biases, limiting the flexibility of model in {Mas} design. \ding{183} \textbf{Agentic Workflow} is also a widely adopted strategy in prior works \cite{zhang2024aflow, zhang2024g, zhuge2024gptswarm, hu2024automated}, aiming at the design of task-level  workflows through an adaptive method. These workflows exhibit limited adaptability across varying in-task queries, which often results in suboptimal trade-offs between performance and cost-efficiency. Therefore, existing methods remain within the realm of semi-autonomous design. 

We argue that the constrained search spaces in recent practices fundamentally restrict the autonomous ability of {Mas}. Candidate pool sampling limits the search space of {Mas} due to predefined structure pool, whereas agentic workflows inherently constrain the {Mas} search to a coarse granularity at the task level. To overcome these limitations, we aim to model the {Mas} construct process over the full-scale graph search space,  enabling fully autonomous and query-adaptive {Mas} design. However, implementing a full-scale graph search to construct autonomous {Mas}  presents significant challenges. The primary challenge stems from the non-Euclidean nature of the {Mas} graph, where the expansive combinatorial space of node feature sampling and edge learning complicates the modeling and optimization process.

\begin{wrapfigure}{r}{0.5\linewidth}
	\centering
	\includegraphics[width=\linewidth]{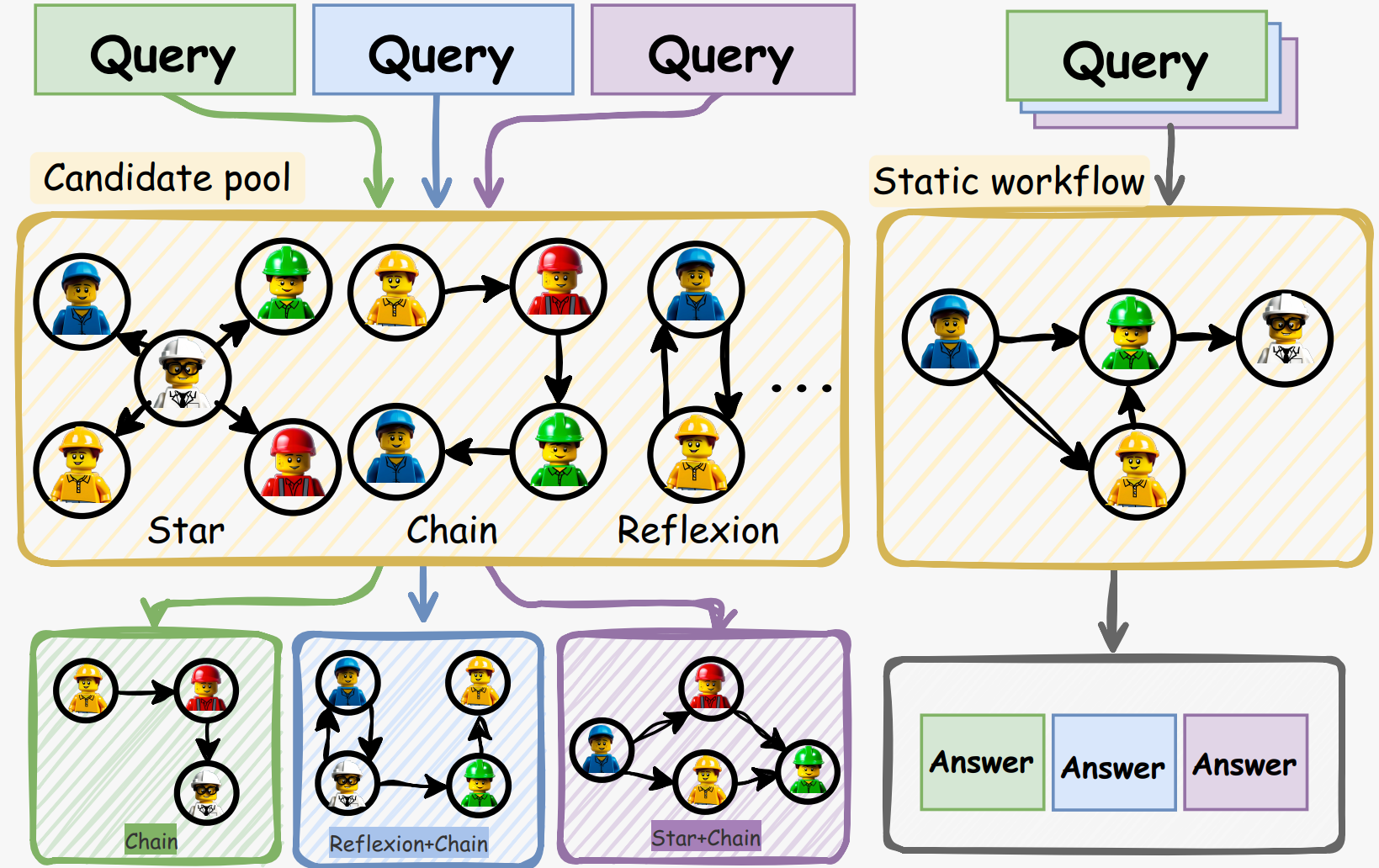}
	\caption{\textbf{\emph{(left)}} Candidate Pool Sampling {Mas}. \textbf{\emph{(right)} } Agentic Workflow.} 
	\vspace{-10pt}
	\label{fig:intro}
\end{wrapfigure}

In this work, we propose an autonomous {Mas} hosting framework (\texttt{MasHost}) based on Reinforcement Learning (RL) algorithm. This design is motivated by the recognition that RL strategy can effectively optimize the exploration of vast search spaces, supported by numerous successful applications \cite{sun2024llm, kaelbling1996reinforcement, li2017deep}. Specifically, we model the design of {Mas} as a graph construction process from scratch under RL guidance. 
\textbf{Firstly}, the challenge lies in the dual-decision nature of the {Mas} construction process, which involves both node role generation and connectivity decision. This differs fundamentally from conventional RL algorithms designed for single-step or sequential actions. Discretizing this dual-action process not only introduces convergence difficulties in high-dimensional combinatorial spaces but also disrupts gradient flow. To address this, we propose a joint probabilistic sampling mechanism that simultaneously models the distribution over agent attributes and their connectivity patterns. Technically, we sample agent roles from the full-scale role space, and subsequently guide the connectivity decisions using joint residual probabilities derived from the role assignments. This mechanism not only ensures efficient representation of the {Mas} design process but also enables the optimization of differentiable sampling. \textbf{Secondly}, the next challenge remains in formulating an effective RL objective that aligns with the autonomous {Mas} construction paradigm. This difficulty arises from the fact that our {Mas} construction is driven by three objectives. Beyond the performance and efficiency goals emphasized in prior {Mas} works, we place additional attention on ensuring the structure rationality of the constructed systems. To achieve this, we propose a novel RL optimization pipeline, Hierarchical Relative Policy Optimization (HRPO), which enables policy-driven {Mas} to respond to queries accurately, efficiently, and rationally.
Inspired by GRPO \cite{shao2024deepseekmath}, HRPO incorporates a hierarchical reward structure that combines group-relative advantages with action-wise absolute rewards. The group-relative advantage strategy compares the relative performance of different {Mas}, guiding the policy network to prioritize accuracy and efficiency in query responses from well-performing {Mas}. The step-wise absolute reward emphasizes the rationality of each action, ensuring that the addition or removal of each agent aligns with the overall objective.
\textbf{Finally}, we conduct comprehensive comparative experiments focusing on three aspects, i.e., performance, cost-efficiency, and rationality. Through empirical comparisons of accuracy and cost-effectiveness with existing state-of-the-art methods, we demonstrate the effectiveness of our \texttt{MasHost}. 
Our \textbf{contributions} can be summarized as:

\begin{itemize}[left=0pt]
	\item We introduce a reinforcement learning-enhanced framework for multi-agent system design, enabling fully autonomous agent generation from scratch.
	\item We propose a joint probabilistic sampling mechanism to realize the dual-action process in Mas construction, along with a hierarchical relative policy optimization  algorithm to optimize the system for high performance, efficiency, and rationality.
	\item Extensive experiments on six benchmarks demonstrate that \texttt{MasHost} consistently outperforms most competitive baselines, validating its effectiveness, efficiency, and structural rationality.
\end{itemize}


\section{Preliminary}
\label{Preliminary}

\subsection{Graph for Multi-agent System}
\label{Preliminary_graph}

Modeling Multi-agent systems ({Mas}) as directed graphs $\mathcal{G} = (\mathcal{V}, \mathcal{E})$ has become a prevailing paradigm in recent researches. Each node $v \in \mathcal{V}$ represents an LLM agent with role-specific attributes that include its capabilities and responsibilities, while each directed edge $e \in \mathcal{E}$ encodes an interaction pathway between agents. This formulation offers a flexible and generalizable abstraction for {Mas}, and recent efforts have advanced this paradigm to design autonomous {Mas} architectures for tackling real-world applications. 



\subsection{Reinforcement Learning for Multi-agent System}
\label{Preliminary_RL}
We formulate the {Mas} construction process as a Markov Decision Process $\mathcal{M} = (\mathcal{S}, \mathcal{A},  \mathcal{R})$. 
\begin{itemize}[left=0pt]
	\item The state $ \mathcal{S}$ covers the global configuration of the {Mas}. At step $t$, the state $s_t \in\mathcal{S}$ encapsulates the  query $Q$, constructed structure $\mathtt{M}_t  = \{ \mathtt{R}_1, ..., \mathtt{R}_{|\mathtt{M}_t|}\}$, and the message list of those agents $\text{MESSAGE}(\mathtt{M}_t)$, i.e., $s_t = \{Q, \mathtt{M}_t, \text{MESSAGE}(\mathtt{M}_t) \}$. Moreover, the output of each agent $\mathtt{R}_j$ can be formalized as $\text{MESSAGE}(\mathtt{R}_j)$, where $j \in[1,|\mathtt{M}_t|]$.
	\item The action space $\mathcal{A}$ defines all possible editing operations for constructing the {Mas} from scratch. It consists of two categories: node-level actions $\mathcal{A}_{n}$ and edge-level actions $\mathcal{A}_{e}$. Specifically, the node-level action ${a}_{n} $ is sampled from $\mathcal{A}_{n} = \{\mathtt{ADD}, \mathtt{DELETE}, \mathtt{EXIT}\}$, corresponding to adding an agent, deleting an agent, and exit the construction process. The edge-level action ${a}_{e} $ include connection sampling operation, denoted as $\mathcal{A}_{e} = \{\mathtt{CONNECT}\}$. Therefore, the atomic action $\mathtt{a}_{t}$ at time step $t$ can be represented as a tuple of two sub-actions, $\mathtt{a}_{t} = ({a}_{n}, {a}_{e})$, corresponding to node-level and edge-level decisions during agent addition.
	\item The policy function $\pi$ governs the  decision-making process of $\mathtt{Mas}$ construction by jointly modeling node-level and edge-level actions. We implement the two levels of actions using two separate parameterized policy networks, denoted as $\pi_{\theta}$ for node-level actions and $\pi_{\phi}$ for edge-level actions.
	\item The reward function ${r}(\mathtt{a}_{t})$ defines the reward of each action $\mathtt{a}_{t} \in \mathcal{A}$ taken in a given state $s \in \mathcal{S}$. To achieve stabilize policy optimization, the advantage function $A(\mathtt{a}_{t})$ is commonly introduced, which  quantifies the relative merit of an action by measuring the difference between the action's expected return and the baseline value of the current state. To this end, the advantage function can be formalized as $A(\mathtt{a}_{t})= Q(\mathtt{a}_{t},s_t)- V(s_t)$, where $Q(\mathtt{a}_{t},s_t)$ is the expected return after taking action $\mathtt{a}_{t}$ in state $s_t$, and $V(s_t)$ is the state-value function representing the expected return from state $s_t$. 
\end{itemize}
Building on the above understanding, the construction of the {Mas} can be formulated within the RL paradigm as a sequence of state-action transitions, represented as $({s_0}, {\mathtt{a}_1}, {s_1}, {\mathtt{a}_2}, {s_2}, \cdots )$, where each state $s_t$ corresponds to the current configuration of the {Mas}, and each action $\mathtt{a}_t$ represents an editing operation that transitions the system from one state to the next.
\subsection{Problem Formulation}
\label{Problem}

Given a query $Q$, this work focuses on leveraging RL to learn an optimal policy $\pi^* =({\pi_{\theta}^*, \pi_{\phi}^*})$ for {Mas} design, enabling fully autonomous and query-specified construction of multi-agent systems. We define the optimality of Mas \texttt{M} from three perspectives: performance quality, resource efficiency, and the structure rationality. Therefore, the overall reward function ${\cal R}(\mathtt{M}\mid Q)$  is formulated as a composition of three key criteria,
\begin{equation}
	{r}(\mathtt{M}\mid Q) = {{r}_{{\rm{perf}}}}(\mathtt{M},Q) + {{r}_{{\rm{eff}}}}(\mathtt{M},Q) + {{r}_{{\rm{struct}}}}({\mathtt{M}}).
	\label{eq:Problem}
\end{equation}
where ${{r}_{{\rm{perf}}}}(\mathtt{M},Q)$ measures performance quality in answering query, ${{r}_{{\rm{eff}}}}(\mathtt{M},Q)$ evaluates resource efficiency in answering query, and ${{r}_{{\rm{struct}}}}({\mathtt{M}})$ captures structure rationality. The objective is to find ${\pi ^*}$ that maximizes the expected reward,
\begin{equation}
	\pi^* = \arg\max_{\pi} \ \mathbb{E}_{\mathtt{M} \sim \pi} \left[ {r}(\mathtt{M} \mid Q) \right].
	\label{eq:Problem_loss}
\end{equation}

\section{Related Work}
\label{Works}

In recent years, the emergence of Large Language Models (LLMs) has introduced new research paradigms for tasks such as mathematical reasoning, code generation, data analysis, and question answering \cite{shao2024deepseekmath, li2024dawn, zhu2024large, xie2024travelplanner, song2023llm, wang2024chain, zha2023tablegpt}. Empirical studies have further shown that challenges unsolved by a single LLM can be effectively addressed through collaborative interactions among multiple LLM-based agents with specialized roles \cite{wei2022chain, yao2023react, shinn2023reflexion}, giving rise to the development of Multi-agent systems (Mas).  Various Mas patterns have been explored, including chain-based, star-shaped, debate-style, and tree-structured frameworks\cite{wei2022chain, zhou2024star, du2023improving, ishibashi2024self, li2024codetree}, leading to notable successes across diverse domains. 

\textbf{Agentic Workflow.} Workflow-based approaches statically perform tasks by following predefined workflows, which is implemented by multiple agents. Designing workflows based on handcraft design and learnable network constitute two prominent application paradigms. The former aims to design workflows based on human understanding and domain knowledge, such as code generation \cite{ridnik2024code}, mathematics \cite{deng2024flow, zhong2024achieving}, and question answering \cite{nori2023can}. The latter focuses on the automated construction of workflows, where an adaptive algorithm  can dynamically design all task-specific workflows. GPTSwarm \cite{zhuge2024gptswarm} models workflows as graphs, and leverages reinforcement learning to design task-specific workflows. ADAS \cite{hu2024automated} represents workflows using code structures and maintains historical workflows in a linear list. AFLOW \cite{zhang2024aflow} also represents workflows through code, emphasizing a custom MCTS algorithm for automated workflow optimization.

\textbf{Autonomous Mas.} Different from workflow-based practices, autonomous Mas efforts focus on designing the most efficient and accurate Mas tailored to each query. MaAS \cite{zhang2025multi} constructs Mas by building an agentic supernet, where each block within the supernet is sampled from a predefined structure pool. MasRouter \cite{yue2025masrouter} constructs Mas by sampling from four structure candidate pools while adaptively learning the number of agents, role types, and LLM types. MAS-GPT \cite{ye2025mas} represents Mas as executable code and trains a LLM to construct Mas by generating code. Actually, existing approaches remain semi-autonomous. The reason lies that most methods model Mas construction as sampling or combining from predefined structure pools. Even for the seemingly fully autonomous framework MAS-GPT, the datasets used to train the LLM are still manually curated rather than generated through exploratory processes. Our work differs fundamentally from existing approaches by employing reinforcement learning to autonomously explore optimal Mas structures from scratch. This design enables the constructed Mas to be free from human biases and solely optimized for better query answering.

\section{\texttt{MasHost}: A Host for Multi-Agent Systems}
\label{Method}

The {Mas} graph serves as a representative example of a non-Euclidean structure. Therefore, the design of {Mas} involves a complex search space that encompasses both node attributes (e.g., agent roles) and connectivity patterns (e.g., inter-agent coordination). As a result, each step of the RL search process exhibits dual-action characteristics. To facilitate efficient search and ensure gradient differentiability, we introduce a Joint Probabilistic Space Sampling (JPSS) mechanism in Sec. \ref{JPSS}. We then analyze the construction objectives in existing {Mas} studies and extend them in our framework from three dimensions. To achieve this goal, we propose a novel Hierarchical Relative Policy Optimization pipeline specifically designed for agent system construction in Sec. \ref{HRPO}.
\begin{figure}
	\centering
	\includegraphics[width=\linewidth]{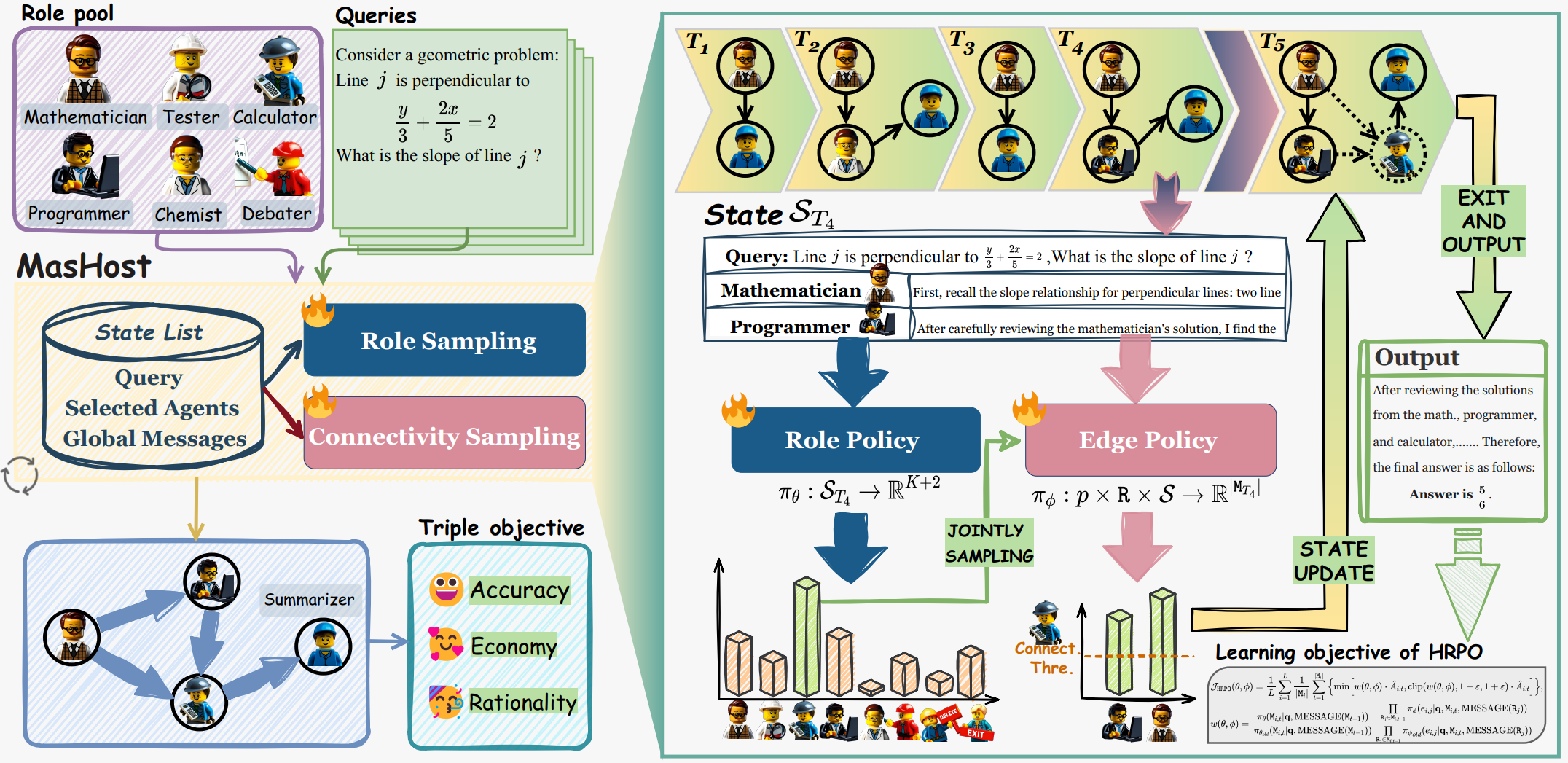}
	\caption{Framework of our \texttt{MasHost}. \textbf{\emph{(left)}} \texttt{MasHost} autonomously manages the complete process of building Mas. \textbf{\emph{(right)}} Detailed construction of Mas using a reinforcement learning strategy.}
	\label{fig:framework}
	\vspace{-10pt}
\end{figure}
 
\subsection{Joint Probability Space Sampling}
\label{JPSS}

The action space $\mathcal{A}$ encompasses all editing operations for constructing {Mas} from scratch, comprising node-level actions $\mathcal{A}_{n}$ and edge-level actions $\mathcal{A}_{e}$. Therefore, the atomic action at time step $t$ is represented as a tuple of two sub-actions, $\mathtt{a}_{t} = ({a}_{n}, {a}_{e})$. Our policy network $\pi $ consists of two components: the first policy ${\pi_\theta}$  selects actions from the space $\mathcal{A}_{n}$, and the secondary policy ${\pi_\phi}$ conducts the link decision from the space $\mathcal{A}_{e}$. 

At the step $t$, the action space $\mathcal{A}_{n}$ is modeled with three types of atomic actions  $ {a}_n \in \{ \mathtt{ADD}, \  \mathtt{DELTE}, \  \mathtt{EXIT}\}$.
\begin{itemize}[left=0pt]
	\item $\mathtt{ADD}$. This action involves adding a new agent. Once triggered, a  agent role is subsequently sampled from the role space $\mathcal{R} = \{ R_1, ..., R_K\} $. Thus, the $\mathtt{ADD}$ action serves both as an activation signal and as a role sapling. In the implementation, we omit its function as an agent-adding signal and instead integrate role selection directly into the policy $\pi_\theta$.  In other words, the single $\mathtt{ADD}$ action is replaced by the role space  $\mathtt{ADD} := \mathcal{R}$.
	\item $\mathtt{DELTE} $. This action corresponds to removing an agent that was most recently modified.
	\item  $\mathtt{EXIT} $. This action marks the completion of {Mas} construction, where the intermediate inference results are passed to a final summary agent, which then produces the answer to the query.
\end{itemize}
Given the above analysis of actions, the sampling space of ${\pi_\theta}$ can be defined as the union of the role space $\mathcal{R}$ and the special actions $\{\texttt{DELETE}, \ \texttt{EXIT}\}$, i.e., $\mathcal{A}_n  = \mathcal{R} \cup \{ \texttt{DELETE}, \ \texttt{EXIT}\}$, where $|\mathcal{A}_n|=K+2$. Given the already constructed {Mas} $\mathtt{M}_{t}$ prior to step $t$, the policy network ${\pi_\theta}$ conducts sampling process with the sate $s_t$ as input, i.e., $ {a}_n \sim  {\pi _\theta }(\mathcal{A}_n |s_t)$. 

Once the sampled action satisfies ${a}_n = \mathtt{R}_t \in \mathcal{R}$, the policy ${\pi_\phi}$ is activated. The policy network ${\pi_\phi}$ is designed to learn the interaction patterns ${a}_e$ between the newly added agent $\mathtt{R}_t$ and the existing agents $\mathtt{M}_t  = \{ \mathtt{R}_1, ..., \mathtt{R}_{|\mathtt{M}_t|}\}$. Technically, ${\pi_\phi}$  performs connectivity sampling using the current state $s_t$ and the selected role $\mathtt{R}_t $ as inputs, $ {a}_e \sim  {\pi _\phi }(\mathcal{A}_e|s_t,\mathtt{R}_t)$.


The independent learning of ${\pi_\theta}$ and ${\pi_\phi}$ is infeasible, which brings the issue of gradient disruption.
To this end, we introduce the JPSS to effectively guide the dual-action decision process in {Mas} design. In the setting of JPSS, the process of constructing a Mas $\texttt{M}$ based on RL is modeled by a unified policy procedure.
Technically, ${\pi_\theta}$ is parameterized to produce a softmax distribution $\mathbb{P}_{a_n}\in \mathbb{R}^{K+2}$ over the role space $\mathcal{A}_n$, and sample the role $\mathtt{R}$ with the highest probability. Subsequently, ${\pi_\phi}$  takes $\mathtt{R}$ as input and outputs a sigmoid-based edge sampling distribution $\mathbb{P}_{a_e} \in \mathbb{R}^{|\mathtt{M}_{t}|}$. Instead of sampling directly from the probability distribution of $\mathbb{P}_{a_e}$, we conduct connectivity sampling based on the joint probability $a_e \sim p \times \mathbb{P}_{a_e}$, where $p$ denotes the probability of selecting $\mathtt{R}$. Under this setup, role selecting and connection learning are modeled as a unified action sampling $\mathtt{a}_t = (a_n, a_e) \sim  {\pi_\theta} \times {\pi_\phi} $, 
\begin{equation}
	{\pi_\theta}: \ \mathcal{S} \to \mathbb{R}^K ,   \  \ \  \  \ {\pi_\phi} : p \times \mathtt{R} \times \mathcal{S}  \to \mathbb{R}^{|\mathtt{M}_{t}|}.
	\label{eq:pi_phi}
\end{equation}

\subsection{Hierarchical Relative Policy Optimization}
\label{HRPO}

We have aligned the {Mas} construction process with the RL by explicitly formulating its atomic policy actions in above discussion. In this subsection, we will introduce the reward mechanism that guides the framework toward learning an optimal {Mas} construction policy.

The evaluation of a {Mas} instance is inherently multi-dimensional, encompassing  its performance quality, resource efficiency, and the rationality of its components. Prior studies have predominantly targeted only one or two of these dimensions, whereas RL enables a unified framework to pursue globally optimal {Mas} across all criteria. To this end, we propose a Hierarchical Relative Optimization (HRPO), which integrates group-relative advantages and step-wise action rewards.

\textbf{Group-relative advantage.} Balancing accuracy and efficiency is the core principle of constructed {Mas}. We introduce an intra-group advantage comparison mechanism to achieve this goal. By comparing relative advantages among instances, this mechanism generates preference signals that drive the policy network to pursue optimal objectives while minimizing resource consumption. 
Specifically, given the initial state $s_0$, we first sample a group of  $\mathtt{Mas}$ instances based on the old policy $\pi_{old}$, denoted as $\mathtt{G} = \{ \mathtt{M}_1, \mathtt{M}_2, \dots, \mathtt{M}_L \}$. Subsequently, instance $\mathtt{M}_i$ is evaluated in terms of both accuracy and resource efficiency in answering the same query $Q$. The reward function ${r_\mathtt{G}}(\cdot)$ is designed as, 
\begin{equation}
	{r_\mathtt{G}}({\mathtt{M}_i}) = \left\{ \begin{array}{l}
		1 - \beta  \cdot {{Tokens}}, \ \ \ \ {\mathtt{M}_i}(Q) = \textbf{Y}\\
		- 1, \qquad \qquad \quad \quad   {\mathtt{M}_i}(Q) \ne \textbf{Y}
	\end{array} \right.
	\label{eq:r_G}
\end{equation}
where $\bf Y$ is the ground-truth of query $Q$ and $\beta$ is a hyper-parameter to ensure $\beta  \cdot {{Tokens}} \in [0, 1]$. Besides, ${Tokens}$ refers to the token usage ( the sum of prompt and completion tokens) by $\mathtt{M}_i$  in answering query $Q$. By implementing reward evaluation on each instance, we can collect the global rewards for the group as $R_\mathtt{G}= \{ {r_\mathtt{G}}({\mathtt{M}_1}), {r_\mathtt{G}}({\mathtt{M}_2}), \dots, {r_\mathtt{G}}({\mathtt{M}_L}) \}$. In order to quantify the policy preferences through comparison, the normalized relative advantage of the $\mathtt{M}_i$ is computed as ${{ A}_\mathtt{G}}(i) = \frac{{{r_\mathtt{G}}(\mathtt{M}_i) - \bar {r_\mathtt{G}}}}{{{\sigma _{r_\mathtt{G}}}}}$, where $\bar{r} = \text{Mean} (R_\mathtt{G})$  and $\sigma_r = \text{Var} (R_\mathtt{G})$. Therefore, ${{ A}_\mathtt{G}} $ distills the strengths and weaknesses of each  $\mathtt{M}$, which can effectively guide the policy network to favor superior patterns during training.

\textbf{Action-wise absolute reward.} Above relative  advantage comparison mechanism can guarantee the performance and efficiency of the {Mas}, but fail to capture the rationality of its internal structure. To this end, we introduce an action-wise absolute reward to explicitly guide the rationality of internal structural design. Early-added agents, which may focus on task decomposition rather than delivering accurate answers, always initially show poor performance. These agents also play a pivotal role in structuring the collaborative process and enabling downstream success. Therefore, it is essential to protect and encourage these early-added agents to ensure the {Mas} fosters reasonable individual collaboration and gradual performance refinement. We introduce an exemption time $\mathcal{T}_{E}$ to safeguard early-stage exploration, where the actions taken before $\mathcal{T}_{E}$ are exempt from penalties, even if they fail to reach the correct solution. Based on this setting, we define the action-wise reward function in as follows:
\begin{equation}
	{r_\mathtt{M_i}}({\mathtt{a}_t}) = \left\{ \begin{array}{l}
		- 1, \qquad \qquad \quad \ \ \text{if} \ \   \mathcal{O}_{t-1}   = \textbf{Y}, \mathcal{O}_{t}   \ne \textbf{Y} \\
		1, \qquad \qquad \qquad \ \text{if} \ \  \mathcal{O}_{t-1}  \ne \textbf{Y}, \mathcal{O}_{t}   = \textbf{Y}\\
		{e^{ - t}}, \qquad \qquad \quad \ \text{if} \ \ \mathcal{O}_{t}= \mathcal{O}_{t-1} = \textbf{Y}  \\
		0, \qquad \qquad \qquad \ \text{if} \ \ t \le \mathcal{T}_{E},\mathcal{O}_{t}= \mathcal{O}_{t-1} \ne \textbf{Y}     \\
		- \alpha  \cdot (t-\mathcal{T}_{E}), \ \ \ \ \text{if} \ \ t > \mathcal{T}_{E}, \mathcal{O}_{t}= \mathcal{O}_{t-1} \ne \textbf{Y}  \\
	\end{array} \right.
	\label{eq:r_a_T}
\end{equation}
where $\alpha$ is a hyper-parameter to ensure $- \alpha  \cdot (t-\mathcal{T}_{E}) \in [-1, 0]$, and $\mathcal{O}_{t-1}$ represents the intermediate output produced by the constructed {Mas} after executing action ${\mathtt{a}_t}$. 
This reward function evaluates the $t$-th action $\mathtt{a}_t$ taken during the construction of $\mathtt{M}_i$, following the principles outlined below.
\begin{itemize}[left=0pt]
	\item $\mathcal{O}_{t-1}   = \textbf{Y}, \mathcal{O}_{t}   \ne \textbf{Y} $. This scenario represents the worst case, where the current action ${\mathtt{a}_t}$ disrupts an already correct {Mas}. Therefore, it should be assigned the maximum penalty, even if it occurs before the exemption time.
	\item $\mathcal{O}_{t-1}  \ne \textbf{Y}, \mathcal{O}_{t}   = \textbf{Y}$. This represents the best-case scenario, indicating that the policy network has successfully captured the correct answering path. To this end, it is assigned the maximum reward when this occurs.
	\item $\mathcal{O}_{t}= \mathcal{O}_{t-1} = \textbf{Y}$. This indicates that consistently correct answers are commendable. However, as the number of exploration steps increases, the reward should decay toward zero.	
	\item  $t \le \mathcal{T}_{E}, \mathcal{O}_{t}= \mathcal{O}_{t-1} \ne \textbf{Y} $. This case indicates that, before the exemption time, the current action ${\mathtt{a}_t}$ neither improves the previous incorrect outcome. This action is neutral and thus free of penalty.
	\item  $t > \mathcal{T}_{E}, \mathcal{O}_{t}= \mathcal{O}_{t-1} \ne \textbf{Y} $. The action ${\mathtt{a}_t}$ fails to bring about any changes in performance after the exemption time. While it does not worsen the result, it is still discouraged. This case may reflect an exploration failure of the policy network. Therefore, a significant penalty $- \alpha  \cdot (t-\mathcal{T}_{E}) \in [-1, 0]$ increasing with $t$ is assigned to this action.	 
\end{itemize}

We have quantified the reward in {Mas} construction from both group-relative preference and action-level reward perspectives. The combination of these hierarchical rewards forms a composite action reward signal that collaboratively guide the policy function to design {Mas} with strong performance, high efficiency, and reasonable components. Building on this hierarchical reward design, the final action advantage ${{\hat A}_{i} (\mathtt{a}_t)}$ for each action ${\mathtt{a}_{t}}$ in ${\mathtt{M}_i}$ is formulated as,
\begin{equation}
	{{\hat A}_{i} (\mathtt{a}_t)} =  {A}_\mathtt{G}(i) +  \sum\limits_{T=t}^{|{\mathtt{M}_i}|} {\gamma ^{T - t}}{r_{\mathtt{M}_i}}({\mathtt{a}_{T}}).
	\label{eq:A_hat}
\end{equation}
The learning objective of our  \texttt{MasHost} based on HPRO policy is formalized by,
\begin{equation}
	\begin{split}
		&{\mathcal{J}_\mathtt{HRPO}}(\theta, \phi)   =  \frac{1}{L}\sum\limits_{i = 1}^{L}   {\frac{1}{{{|\mathtt{M}_i|}}}\sum\limits_{t = 1}^{|\mathtt{M}_i|} {\left\{ {\min \left[ {w(\theta, \phi) \cdot {{\hat A}_{i,t}},  \text{clip} (w(\theta, \phi),1 - \varepsilon ,1 + \varepsilon ) \cdot {{\hat A}_{i,t}}} \right] } \right\}} }, \\
		&   w(\theta, \phi) =  \frac{{{\pi _\theta }({\mathtt{M}_{i,t}}|\textbf{q},{{\text{MESSAGE}({\mathtt{M}_{i,t-1}})}})}}{{{\pi _{{\theta _{old}}}}({\mathtt{M}_{i,t}}|\textbf{q},{\text{MESSAGE}({\mathtt{M}_{i,t-1}})})} }
		\frac{\prod\limits_{\mathtt{R}_{{j}} \in \mathtt{M}_{i,{t-1}}} {{\pi _\phi }({e_{i,j}}|\textbf{q},{\mathtt{M}_{i,t-1}},\text{MESSAGE}(\mathtt{R}_{{j}}))} }{ \prod\limits_{\mathtt{R}_{{j}} \in \mathtt{M}_{i,{t-1}}} {{{\pi _\phi }_{old}}({e_{i,j}}|\textbf{q},{\mathtt{M}_{i,t-1}},\text{MESSAGE}(\mathtt{R}_{{j}}))} },
	\end{split}
	\label{eq:loss}
\end{equation}
where ${\pi_\theta}$ and ${\pi_\phi}$ denote the current policy models, and ${{\pi_\theta}_{{old}}}$ and ${{\pi_\phi}_{{old}}}$ represent the corresponding old policy models. $ \varepsilon$ is a clipping-related hyper-parameter  introduced in PPO \cite{schulman2017proximal} for stabilizing training.  Similarly, $w(\theta, \phi)$ denotes the importance sampling ratio, also introduced in PPO, which serves to constrain excessive policy updates by adjusting the weight of sampled $\mathtt{Mas}$.

\section{Autonomy and Rationality Guarantee}
\label{Guarantee}

We guarantee the autonomous capability of \texttt{MasHost} to construct multiple agents from two complementary perspectives. Our HRPO-based graph growth mechanism can generate arbitrary directed graphs, while our role sampling strategy, in contrast to prior methods restricted to task-specific role pools, operates over the entire role space.

\textbf{Autonomy in graph construction.} From a graph-theoretic perspective, we argue that the design space explored by \texttt{MasHost} is equivalent to the entire set of directed graphs over a given node set. Specifically, by modeling node role assignments and edge connectivity as joint probabilistic variables, our framework ensures the representational completeness of all possible Mas interaction topologies without structural bias or limitation. This guarantee implies that \texttt{MasHost} can generate any feasible directed graph configuration, thus achieving full autonomy in graph construction.

\textbf{Autonomy of role selection.} The autonomous capability of role selection is largely overlooked in existing works, which typically preset a task-specific role pool and select agent roles within this limited space. In this work, we focus on enabling autonomous role selection by sampling from the entire role space without human-imposed restrictions. This approach not only enhances the flexibility and generality of the system but also allows for emergent agent behaviors that are better aligned with dynamic task demands. To address the associated optimization challenges arising from the high-dimensional and combinatorial nature of the full role space, we introduce a joint probabilistic modeling framework that guides role sampling in a stable and differentiable manner.

\begin{algorithm}[ht]
	\caption{\textsc{\texttt{MasHost}}: RL-based Multi-Agent System Construction}
	\begin{algorithmic}[1]
		\Require Query $Q$, full-scale role pool $\mathcal{R}$
		\Ensure Multi-agent System Graph $\mathtt{M}$
		\State Initialize policy networks $\pi_\theta$ (node-level), $\pi_\phi$ (edge-level)
		\State Initialize empty MAS graph $\mathtt{M} \leftarrow \emptyset$, $s_0=\{Q\}$
		\While{not \textsc{Terminated}($\mathtt{M}$)}
		\State Observe current state $s_t = \{Q, \mathtt{M}, \textsc{Message}(\mathtt{M})\}$
		\State Sample 4 cases to construct a relative group $\mathtt{G} = \{ \mathtt{M}_1, \mathtt{M}_2, \mathtt{M}_3, \mathtt{M}_4 \} \sim \pi$
		\State Sample action $a_n \sim \pi_\theta(a_n \mid s_t)$ \Comment{Node-level action}
		\If{$a_n = \textsc{EXIT}$}
		\State \textbf{break}
		\ElsIf{$a_n = \textsc{DELETE}$}
		\State Remove last-added agent from $\mathtt{M}$
		\Else
		\State Add agent $v$ with role $a_n$ to $\mathtt{M}$
		\State Sample edge distribution $P_e \leftarrow \pi_\phi(a_e \mid s_t, a_n)$ \Comment{Edge-level action}
		\State Sample connections $a_e \sim p(a_n) \cdot P_e$ \Comment{Joint distribution sampling}
		\State Add edges $a_e$ to $\mathtt{M}$
		\EndIf
		\State Compute group-relative preference  ${A}_\mathtt{G}(i)$ and action-level reward ${r_{\mathtt{M}_i}}({\mathtt{a}_{T}})$
		\State Compute advantage ${{\hat A}_{i} (\mathtt{a}_t)} =  {A}_\mathtt{G}(i) +  \sum\limits_{T=t}^{|{\mathtt{M}_i}|} {\gamma ^{T - t}}{r_{\mathtt{M}_i}}({\mathtt{a}_{T}})$
		\State Update $\pi_\theta, \pi_\phi$ via HRPO objective ${\mathcal{J}_\mathtt{HRPO}}(\theta, \phi) $
		\EndWhile
		\State \Return $\mathtt{M}$
	\end{algorithmic}
\end{algorithm}

\definecolor{D88B99}{HTML}{841F27}
\definecolor{90CD85}{HTML}{00B050}
\definecolor{FFE5E5}{HTML}{FFE5E5}
\begin{table*}[t]
	\centering
	\caption{Performance comparison with single agent execution methods, hand-craft multi-agent systems, agentic workflows, and autonomous mutli-agent systems. The execution LLM is consistently set as {gpt-4o-mini} for all baselines.We report the average performance across five independent runs.
	}
	\label{tab:results}
	\resizebox{\textwidth}{!}{
		\begin{tabular}{lcccccc>{\columncolor{FFE5E5}}c}
			\toprule
			\textbf{Methods} & \textbf{GSM8K} & \textbf{MATH} & \textbf{MMLU} & \textbf{ GPQA } & \textbf{MBPP} & \textbf{HumanEval} & \cellcolor{FFE5E5}\textbf{Average} \\ \hline 
			IO \cite{openai2024gpt4omini} &
			87.37 & 46.32 & 81.53 & 39.21 & 71.62 & 87.21 & \cellcolor{FFE5E5}68.88 \\
			CoT \cite{wei2022chain} &
			86.85\textsubscript{\textcolor{90CD85}{$\downarrow$0.52}} & 45.83\textsubscript{\textcolor{90CD85}{$\downarrow$0.49}} & 81.92\textsubscript{\textcolor{D88B99}{$\uparrow$0.39}} & 39.20\textsubscript{\textcolor{90CD85}{$\downarrow$0.01}} & 71.21\textsubscript{\textcolor{90CD85}{$\downarrow$0.41}} & 88.39\textsubscript{\textcolor{D88B99}{$\uparrow$1.18}} & \cellcolor{FFE5E5}68.90 \\
			SC (CoT$\times$5) \cite{wang2022self} &
			87.86\textsubscript{\textcolor{D88B99}{$\uparrow$0.49}} & 47.79\textsubscript{\textcolor{D88B99}{$\uparrow$1.47}} & 80.65\textsubscript{\textcolor{90CD85}{$\downarrow$0.88}} & 38.98\textsubscript{\textcolor{90CD85}{$\downarrow$0.23}} & 72.87\textsubscript{\textcolor{D88B99}{$\uparrow$1.25}} & 88.37\textsubscript{\textcolor{D88B99}{$\uparrow$1.16}} & \cellcolor{FFE5E5}69.42 \\ \hline
			MultiPersona \cite{wang2023unleashing} &
			87.12\textsubscript{\textcolor{90CD85}{$\downarrow$0.25}} & 43.97\textsubscript{\textcolor{90CD85}{$\downarrow$2.35}} & 81.03\textsubscript{\textcolor{90CD85}{$\downarrow$0.50}} & 40.09\textsubscript{\textcolor{D88B99}{$\uparrow$0.88}} & 72.18\textsubscript{\textcolor{D88B99}{$\uparrow$0.56}} & 87.54\textsubscript{\textcolor{D88B99}{$\uparrow$0.33}} & \cellcolor{FFE5E5}68.66 \\
			LLM-Debate \cite{du2023improving}  &
			88.52\textsubscript{\textcolor{D88B99}{$\uparrow$1.15}} & 47.33\textsubscript{\textcolor{D88B99}{$\uparrow$1.01}} & 82.44\textsubscript{\textcolor{D88B99}{$\uparrow$0.91}} & 39.57\textsubscript{\textcolor{D88B99}{$\uparrow$0.36}} & 69.82\textsubscript{\textcolor{90CD85}{$\downarrow$1.80}} & 88.07\textsubscript{\textcolor{D88B99}{$\uparrow$0.86}} & \cellcolor{FFE5E5}69.29 \\
			DyLAN \cite{liu2023dynamic} &
			89.21\textsubscript{\textcolor{D88B99}{$\uparrow$1.84}} & 48.19\textsubscript{\textcolor{D88B99}{$\uparrow$1.87}} & 81.90\textsubscript{\textcolor{D88B99}{$\uparrow$0.37}} & 40.54\textsubscript{\textcolor{D88B99}{$\uparrow$1.33}} & 76.50\textsubscript{\textcolor{D88B99}{$\uparrow$4.88}} & 86.98\textsubscript{\textcolor{90CD85}{$\downarrow$0.23}} & \cellcolor{FFE5E5}70.55 \\ \hline
			GPTSwarm \cite{zhuge2024gptswarm}  &
			88.34\textsubscript{\textcolor{D88B99}{$\uparrow$0.97}} & 48.31\textsubscript{\textcolor{D88B99}{$\uparrow$1.99}} & 81.49\textsubscript{\textcolor{90CD85}{$\downarrow$0.04}} & 42.41\textsubscript{\textcolor{D88B99}{$\uparrow$3.20}} & 77.34\textsubscript{\textcolor{D88B99}{$\uparrow$5.72}} & 88.29\textsubscript{\textcolor{D88B99}{$\uparrow$1.08}} & \cellcolor{FFE5E5}71.03 \\
			ADAS \cite{hu2024automated}  &
			85.72\textsubscript{\textcolor{90CD85}{$\downarrow$1.65}} & 41.70\textsubscript{\textcolor{90CD85}{$\downarrow$4.62}} & 80.61\textsubscript{\textcolor{90CD85}{$\downarrow$0.92}} & 39.80\textsubscript{\textcolor{D88B99}{$\uparrow$0.59}} & 68.00\textsubscript{\textcolor{90CD85}{$\downarrow$3.62}} & 83.79\textsubscript{\textcolor{90CD85}{$\downarrow$3.42}} & \cellcolor{FFE5E5}66.60 \\
			AFlow \cite{zhang2024aflow}   &
			90.60\textsubscript{\textcolor{D88B99}{$\uparrow$3.23}} & 50.63\textsubscript{\textcolor{D88B99}{$\uparrow$4.31}} & 81.93\textsubscript{\textcolor{D88B99}{$\uparrow$0.40}} & 44.23\textsubscript{\textcolor{D88B99}{$\uparrow$5.02}} & 80.94\textsubscript{\textcolor{D88B99}{$\uparrow$9.32}} & 89.27\textsubscript{\textcolor{D88B99}{$\uparrow$2.06}} & \cellcolor{FFE5E5}72.94\\ \hline
			AutoAgents \cite{chen2023autoagents} &
			87.36\textsubscript{\textcolor{90CD85}{$\downarrow$0.01}} & 43.94\textsubscript{\textcolor{90CD85}{$\downarrow$2.38}} & 82.00\textsubscript{\textcolor{D88B99}{$\uparrow$0.47}} & 42.57\textsubscript{\textcolor{D88B99}{$\uparrow$3.36}} & 71.11\textsubscript{\textcolor{90CD85}{$\downarrow$0.51}} & 86.95\textsubscript{\textcolor{90CD85}{$\downarrow$0.26}} & \cellcolor{FFE5E5}68.99 \\
			MAS-GPT \cite{ye2025mas} &
			91.36\textsubscript{\textcolor{D88B99}{$\uparrow$3.99}} & 52.11\textsubscript{\textcolor{D88B99}{$\uparrow$5.79}} & 82.09\textsubscript{\textcolor{D88B99}{$\uparrow$0.56}} & 44.91\textsubscript{\textcolor{D88B99}{$\uparrow$5.70}} & 80.19\textsubscript{\textcolor{D88B99}{$\uparrow$8.57}} & 87.76\textsubscript{\textcolor{D88B99}{$\uparrow$0.55}} & \cellcolor{FFE5E5}73.07 \\
			G-Designer \cite{zhang2024g}  &
			91.27\textsubscript{\textcolor{D88B99}{$\uparrow$3.90}} & 50.03\textsubscript{\textcolor{D88B99}{$\uparrow$3.71}} & 81.44\textsubscript{\textcolor{90CD85}{$\downarrow$0.09}} & 42.02\textsubscript{\textcolor{D88B99}{$\uparrow$2.81}} & 80.10\textsubscript{\textcolor{D88B99}{$\uparrow$8.48}} & 87.32\textsubscript{\textcolor{D88B99}{$\uparrow$0.11}} & \cellcolor{FFE5E5}72.03 \\
			MaAS \cite{zhang2025multi}  &
			91.76\textsubscript{\textcolor{D88B99}{$\uparrow$4.39}} & 51.71\textsubscript{\textcolor{D88B99}{$\uparrow$4.40}} & 83.17\textsubscript{\textcolor{D88B99}{$\uparrow$1.64}} & 44.39\textsubscript{\textcolor{D88B99}{$\uparrow$5.18}} & 80.21\textsubscript{\textcolor{D88B99}{$\uparrow$8.59}} & 90.09\textsubscript{\textcolor{D88B99}{$\uparrow$2.88}} & \cellcolor{FFE5E5} \underline{73.56} \\ \hline
			\textbf{\texttt{MasHost} } (Ours)&
			{93.23}\textsubscript{\textcolor{D88B99}{$\uparrow$5.86}} & {52.42}\textsubscript{\textcolor{D88B99}{$\uparrow$6.10}} & {83.40}\textsubscript{\textcolor{D88B99}{$\uparrow$1.87}} & {45.19}\textsubscript{\textcolor{D88B99}{$\uparrow$5.98}} & {80.97}\textsubscript{\textcolor{D88B99}{$\uparrow$9.35}} & {89.96}\textsubscript{\textcolor{D88B99}{$\uparrow$2.75}} & \cellcolor{FFE5E5} \textbf{74.20} \\
			\bottomrule
		\end{tabular}
		}
		\vspace{-10pt}
\end{table*}

\section{Experiments}
\label{Experiments}

\subsection{Experimental Setup}
\label{setup}
\textbf{Datasets.} We evaluate our \texttt{MasHost} on six widely-used public benchmarks, including (1) math reasoning: {GSM8K} \cite{cobbe2021training}, {MATH} \cite{hendrycks2021measuring}; (2) question-answering: {GPQA} \cite{rein2024gpqa}, MMLU \cite{hendrycks2020measuring}; (3) code generation: {HumanEval} \cite{chen2021evaluating}, {MBPP} \cite{austin2021program}.

\textbf{Baselines.} We compare mutli-agent systems constructed by \texttt{MasHost} against various types of baselines, including (1) single agent execution methods: IO \cite{openai2024gpt4omini}, Chain-of-Thought (CoT) \cite{wei2022chain}, CoT SC (5-shot) \cite{wang2022self}; (2) hand-craft multiagent systems: MultiPersona \cite{wang2023unleashing}, LLM-Debate \cite{du2023improving}, DyLAN \cite{liu2023dynamic}; (3) agentic workflows: GPTSwarm \cite{zhuge2024gptswarm}, ADAS \cite{hu2024automated}, AFlow \cite{zhang2024aflow}; (4) autonomous mutli-agent systems: AutoAgents \cite{chen2023autoagents}, MAS-GPT \cite{ye2025mas}, G-Designer \cite{zhang2024g}, MaAS \cite{zhang2025multi}.

\textbf{Implementation Details.}  Following the experimental settings adopted by most baselines \cite{zhang2024aflow, zhang2025multi}, we select  GPT-4o-mini-0718 \cite{openai2024gpt4omini} as the LLM executor, which is accessed via APIs. Besides, we set the temperature to 0 for the executor. We implement our \texttt{MasHost} on a server equipped with an NVIDIA A100-SXM4-80GB GPU. 

\textbf{Metrics.} For GSM8K, MATH, GPQA and MMLU, we report the Accuracy (\%) as the metric. For HumanEval and MBPP, we report the Pass@1 (\%)  to assess code accuracy. 

\subsection{Performance Comparison}
\label{Performance}

\begin{figure}[ht]
	\centering
	\begin{subfigure}[b]{0.28\textwidth}
		\centering
		\includegraphics[width=\textwidth]{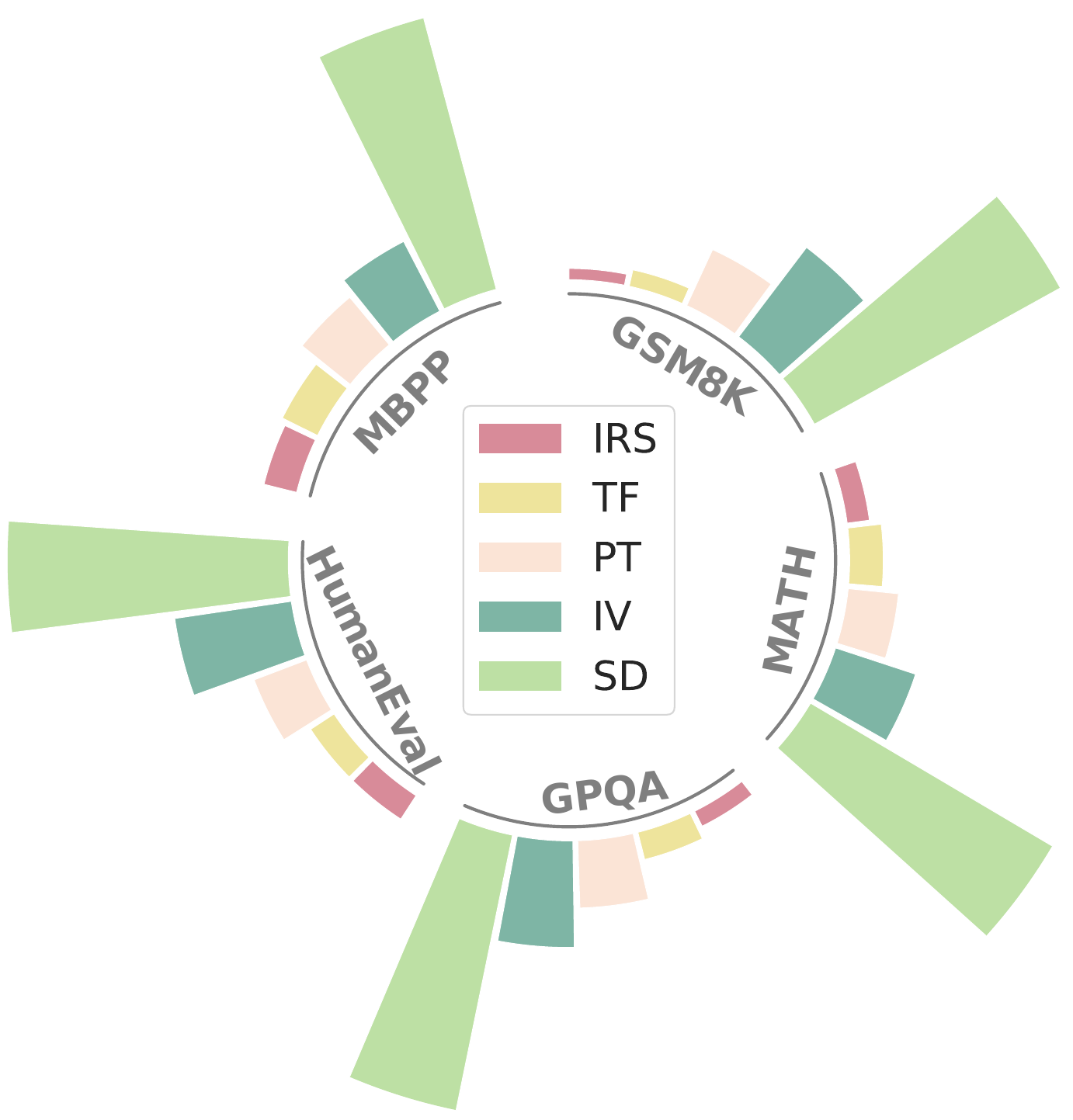}
		\label{fig:Rationality_M_1}
	\end{subfigure}
	\begin{subfigure}[b]{0.30\textwidth}
		\centering
		\includegraphics[width=\textwidth]{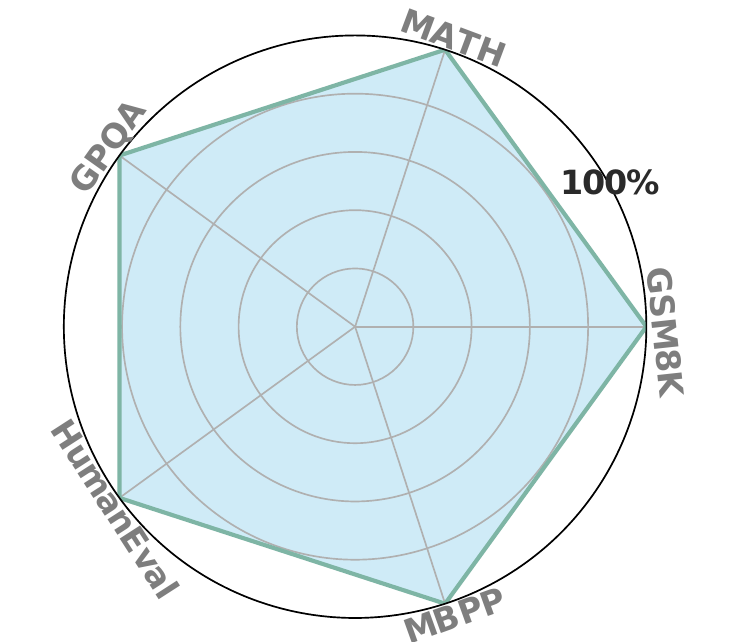}
		\label{fig:Rationality_M_3}
	\end{subfigure}
	\begin{subfigure}[b]{0.345\textwidth}
		\centering
		\includegraphics[width=\textwidth]{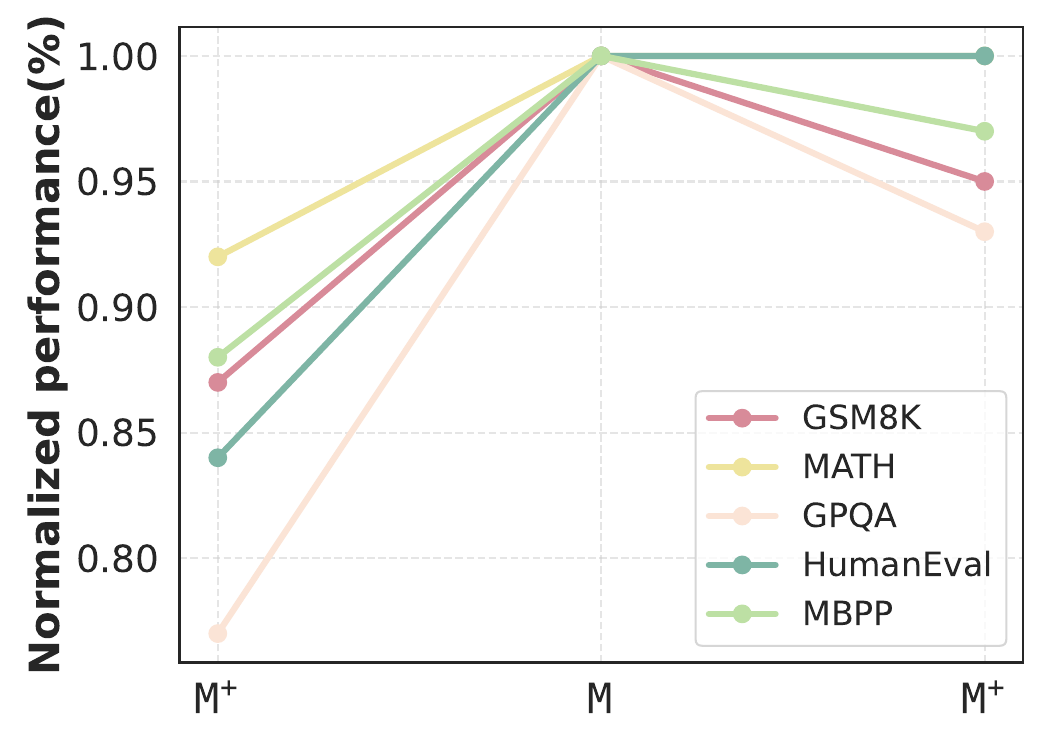}
		\label{fig:Rationality_M_4}
	\end{subfigure}
	\caption{\textbf{\emph{(left)}} Robustness of \texttt{MasHost}. \textbf{\emph{(middle)} } The similarity between the roles and the queries type. \textbf{\emph{(right)}}  Rationality of Constructed Mas.}
	\label{fig:Rationality_2}
\end{figure}

As shown in Tab. \ref{tab:results}, our proposed \texttt{MasHost} consistently achieves the best performance among all compared methods. Compared to the existing state-of-the-art, our \texttt{MasHost} achieves an absolute performance improvement of up to 1.47 \% on the GSM8k, highlighting its superiority over existing methods. Furthermore, we also focus on the samples where \texttt{MasHost} failed to provide correct answers to further investigate its robustness. We categorize the samples with incorrect answers into five types: 
(1) global failure due to \emph{Incorrect Role Selection} (\emph{IRS}), (2) target omission caused by \emph{Task Forgetting} (\emph{TF}), (3) incomplete answers caused by \emph{Premature Termination} (\emph{PT}), (4) \emph{Incorrect Verification} (\emph{IV}), and (5) correct reasoning with \emph{Slight Deviations} in the final result (\emph{SD}). As shown in Fig. \ref{fig:Rationality_2}\emph{(left)}, we observe that the erroneous samples are primarily concentrated in two categories: IV and SD. This indicates that \texttt{MasHost} is able to identify the correct direction for answering but fails to produce the correct solution due to the complexity and difficulty of the questions. This demonstrates the potential of \texttt{MasHost} in tackling complex problems and highlights its robustness.

\subsection{Cost-efficient Analysis}
\label{Cost}
\begin{table}  
	\centering
	\caption{\small Efficiency comparison on the MATH Benchmark. Best results are in \textbf{bold}.}
	\label{tab:efficiency_comparison}
	\small
	\begin{tabular}{@{}lccc@{}}
		\toprule
		\textbf{Methods} & \shortstack{ \# Prompt \\ Tokens} & \shortstack{ \# Completion \\ Tokens} & \shortstack{Cost \\ ($\times 10^{-3}$~\emph{USD})} \\
		\midrule
		GPTSwarm & {6,322} & \textbf{3,379} & 2.976 \\
		AFlow    & 4,938 & 3,943 & 3.106 \\
		MaAS     & 5,273 & 3,749 & 3.040 \\ \hline
		\textbf{\texttt{MasHost}} & \textbf{3,630} & {3,698} & \textbf{2.763} \\
		\bottomrule
	\end{tabular}
\end{table}
As shown in Tab. \ref{tab:efficiency_comparison}, we present the average cost required to answer each query in the test phase, using GPT-4o-mini as execution LLM. The cost efficiency of our \texttt{MasHost} is highly competitive. Actually, we have incorporated the following design strategies into our framework to reduce costs. (1) The inter-group advantage in HRPO takes cost consumption into account and quantifies the associated loss. (2) The global message pool prevents redundant invocations of the same role. Therefore, we conclude that our \texttt{MasHost} provides performance improvements while maintaining cost efficiency.

\subsection{Rationality Discussion}
\label{Rationality}

We assess the rationality of the multi-agent system built by \texttt{MasHost} from two aspects: (1) the role rationality and (2) the structure rationality.

\textbf{Rationality of role assignment.}  Given the full-scale role space search in our work, ensuring the rationality of role selection is essential for tackling complex real-world queries. We design a correlation matching strategy to verify whether each role in the constructed Mas is relevant to the given query. As shown in Fig. \ref{fig:Rationality_2}\emph{(middle)}, we observe a perfect correlation (i.e., 100\%) between the assigned roles and query types across all datasets. This demonstrates that even under full-space role search, the multi-agent system constructed by \texttt{MasHost} maintains explainable rationality.  

\textbf{Rationality of Mas structure.} We evaluate the rationality of our Mas structure in terms of redundancy and oversimplification. Let \texttt{M} denote the multi-agent system generated by \texttt{MasHost}, where removing one agent yields $\mathtt{M^{-}}$ and adding one task-related agent results in $\mathtt{M^{+}}$. We sample 100 instances from each of the GSM8K and HumanEval datasets to compare the performance of \texttt{M}, $\mathtt{M^{-}}$, and $\mathtt{M^{+}}$, thereby verifying the rationality of the constructed MAS. As shown in Fig \ref{fig:Rationality_2}\emph{(fight)}, we observe that, compared to \texttt{M}, the performance of $\mathtt{M^{-}}$ exhibits a significant drop, while the performance of $\mathtt{M^{+}}$ show a slight performance degradation.  The decline in performance resulting from the addition of agents is primarily due to incorrect post-processing, which can corrupt previously accurate information. This indicates that the \texttt{M} constructed by \texttt{MasHost} achieves an efficient, accurate, and reasonable multi-agent system.

\definecolor{mycellcolor}{HTML}{D88B99}

\begin{table}
	\centering
	\caption{Ablation study of \texttt{MasHost}. \emph{Cost} refers to the relative proportion of total token consumption during the training, with \texttt{MasHost} normalized to $1.00$.}
	\label{tab:ablation}
	\begin{tabular}{llcc|cc}
		\toprule
		\multirow{2}{*}{\textbf{Dataset}} &  & \multicolumn{2}{c|}{\textbf{HumanEval}} & \multicolumn{2}{c}{\textbf{GSM8K}} \\
		\cmidrule(lr){3-4} \cmidrule(lr){5-6}
		& & Perf. & Cost & Perf. & Cost \\
		\midrule
		\multicolumn{2}{l}{\texttt{MasHost}} & 89.96 & 1.00 & 93.23 & 1.00 \\
		\midrule
		\multicolumn{2}{l}{\texttt{MasHost} \emph{w.o. JPSS}} & 88.07 & 1.10 & 91.53 & 1.02 \\
		\multicolumn{2}{l}{\texttt{MasHost} \emph{w.o. HRPO}} & 87.22 & 1.43 &  90.64 & 1.73 \\
		\multicolumn{2}{l}{\texttt{MasHost} \emph{w.o. ET}} & 88.93 & 0.92 & 91.17 & 0.96 \\
		\bottomrule
	\end{tabular}
\end{table}

\subsection{Ablation Study}
\label{Ablation}
We conduct ablation studies to explore the effectiveness of each component of \texttt{MasHost}. Specifically, we analyze the respective impacts of three core components: the joint probabilistic space sampling mechanism (\emph{JPSS}), hierarchical relative policy optimization (\emph{HRPO}), and the design of exemption time (\emph{ET}). To this end, we design three variants based on \texttt{MasHost}, \texttt{MasHost} \emph{w.o. JPSS}, MasHost \emph{w.o. HRPO}, and MasHost \emph{w.o. ET}. 
Tab. \ref{tab:ablation} shows that the performance drops significantly when any of the three core components is removed. Among them, \texttt{MasHost} \emph{w.o. HRPO} exhibits the most significant performance drop, indicating that this component has the greatest impact on performance. Although \texttt{MasHost} \emph{w.o. ET} has a relatively smaller effect on performance, the resulting multi-agent systems often converge to a smaller scale. In this case, many of the resulting structures lack rationality and fail to handle complex tasks effectively.

\begin{figure}
	\centering
	\begin{subfigure}[b]{0.4\textwidth}
		\centering
		\includegraphics[width=\textwidth]{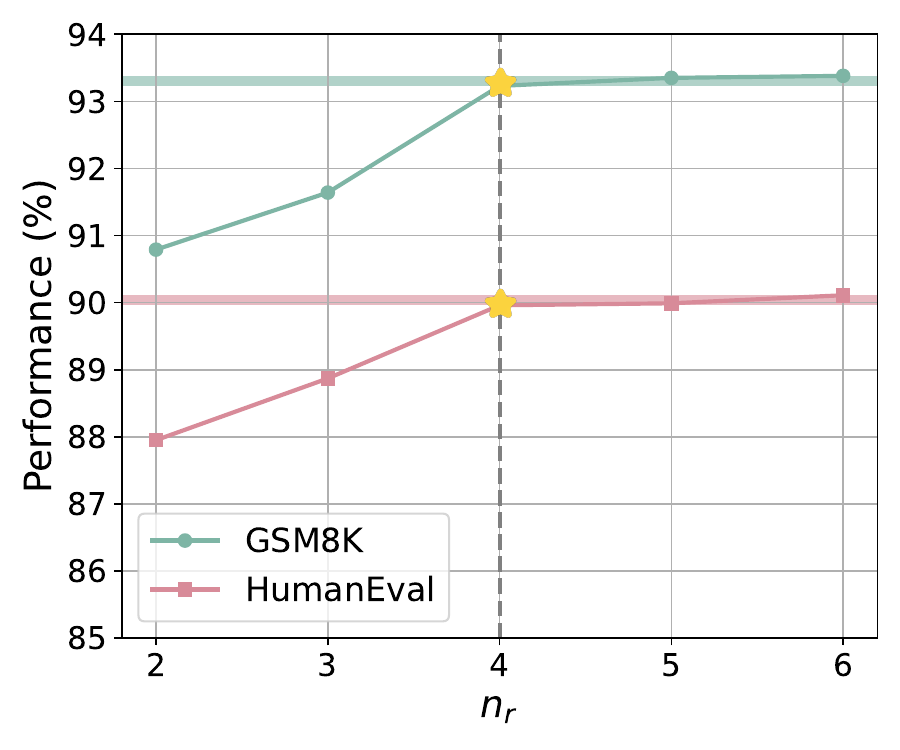}
		\label{fig:Sensitivity_1}
	\end{subfigure}
	\begin{subfigure}[b]{0.4\textwidth}
		\centering
		\includegraphics[width=\textwidth]{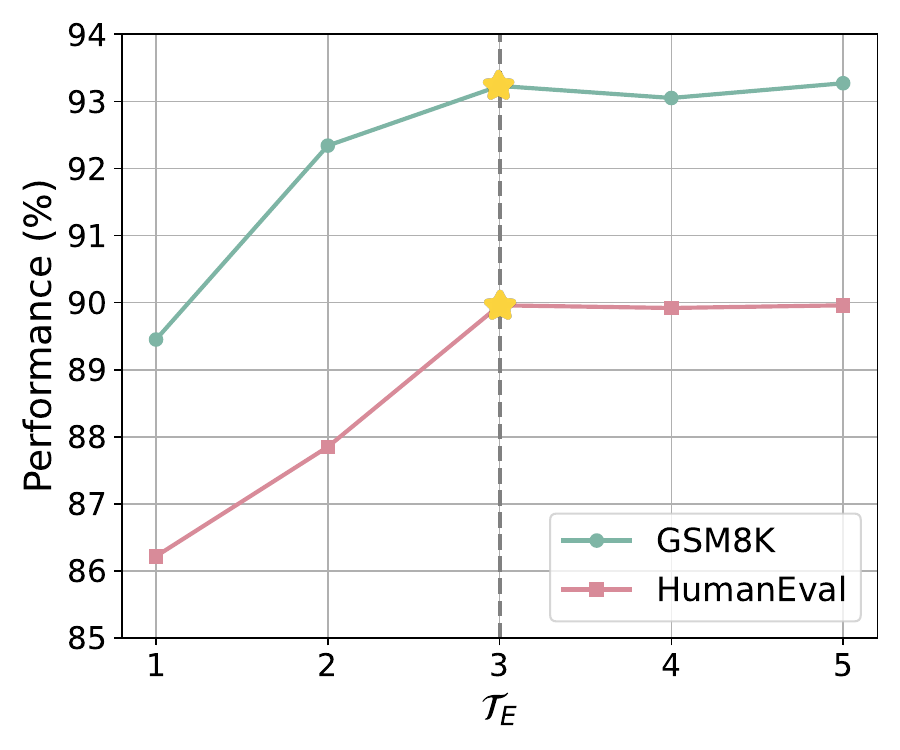}
		\label{fig:Sensitivity_2}
	\end{subfigure}
	\caption{The sensitivity of training rounds $n_r$, and exemption time $\mathcal{T}_{E}$.}
	\label{fig:Sensitivity}
\end{figure}
\subsection{Sensitivity Analysis}
\label{Sensitivity}

We investigate the sensitivity of training rounds $n_r$, and exemption time $\mathcal{T}_{E}$. As shown in Fig. \ref{fig:Sensitivity}, we present the performance fluctuations under different hyper-parameter settings on  GSM8K and HumanEval. Although performance improves with larger $n_r$, the marginal gains diminish when $n_r > 4$. Therefore, we fix $n_r=4$ to achieve a trade-off between performance and cost. We observe that the performance converges once the  exemption time  $\mathcal{T}_{E}> 3$. Given that the value of  $\mathcal{T}_{E}$ is proportional to the cost consumption, we set $\mathcal{T}_{E}=3$. 

\subsection{Visualization Results}
\label{Visualization}

\begin{figure}
	\centering
	\includegraphics[width=\linewidth]{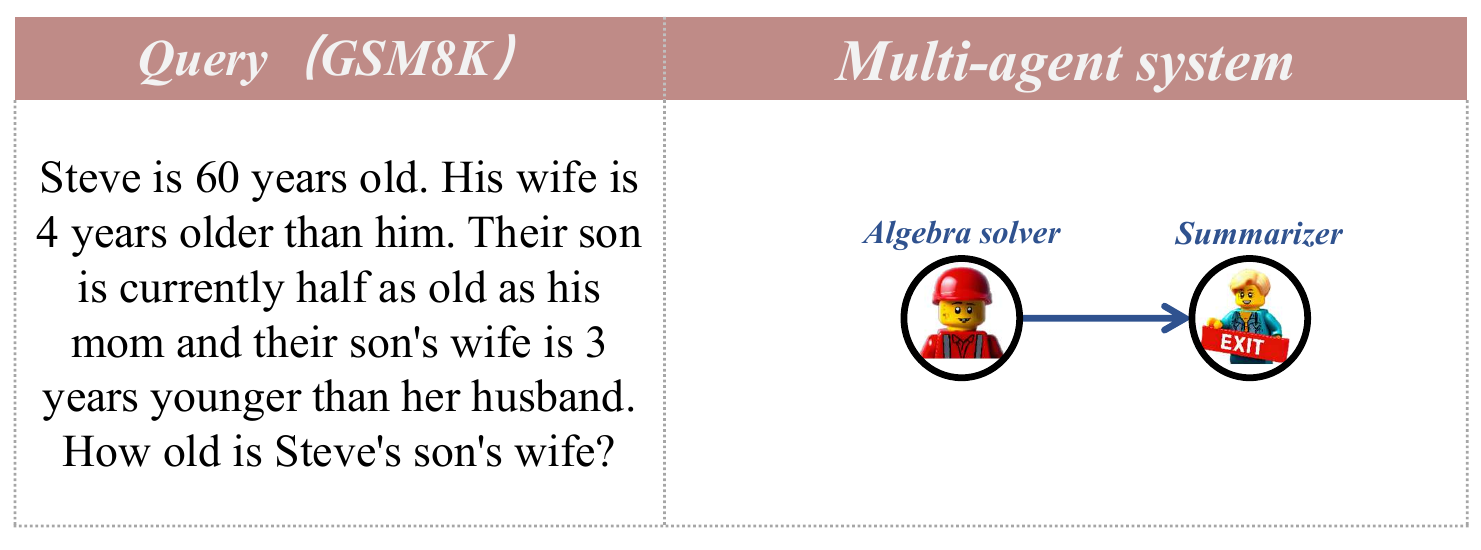}
	\caption{The Mas constructed on the GSM8K sample.} 
	\label{fig:demo_GSM8K}
\end{figure}

\begin{figure}
	\centering
	\includegraphics[width=\linewidth]{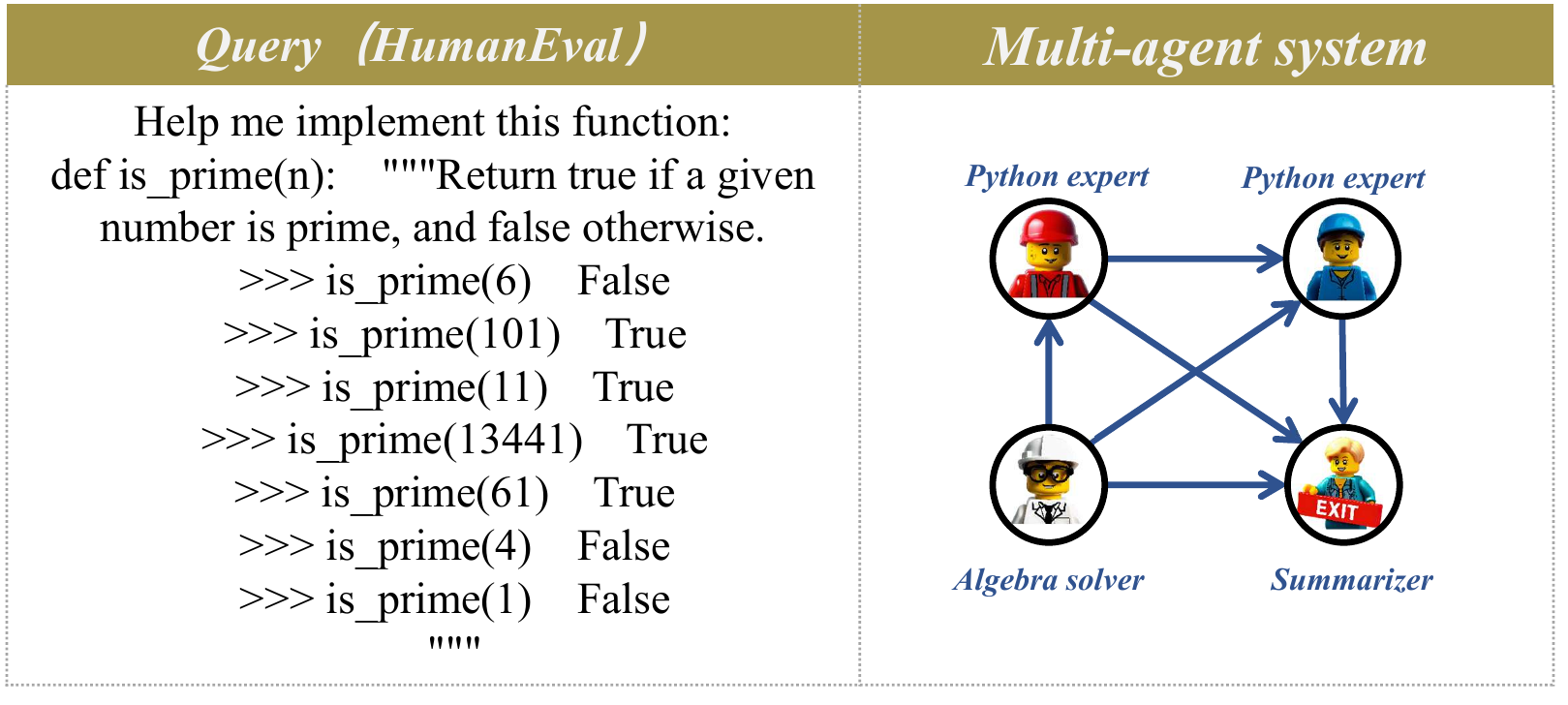}
	\caption{The Mas constructed on the HumanEval sample.} 
	\label{fig:demo_HumanEval}
\end{figure}

\begin{figure}
	\centering
	\includegraphics[width=\linewidth]{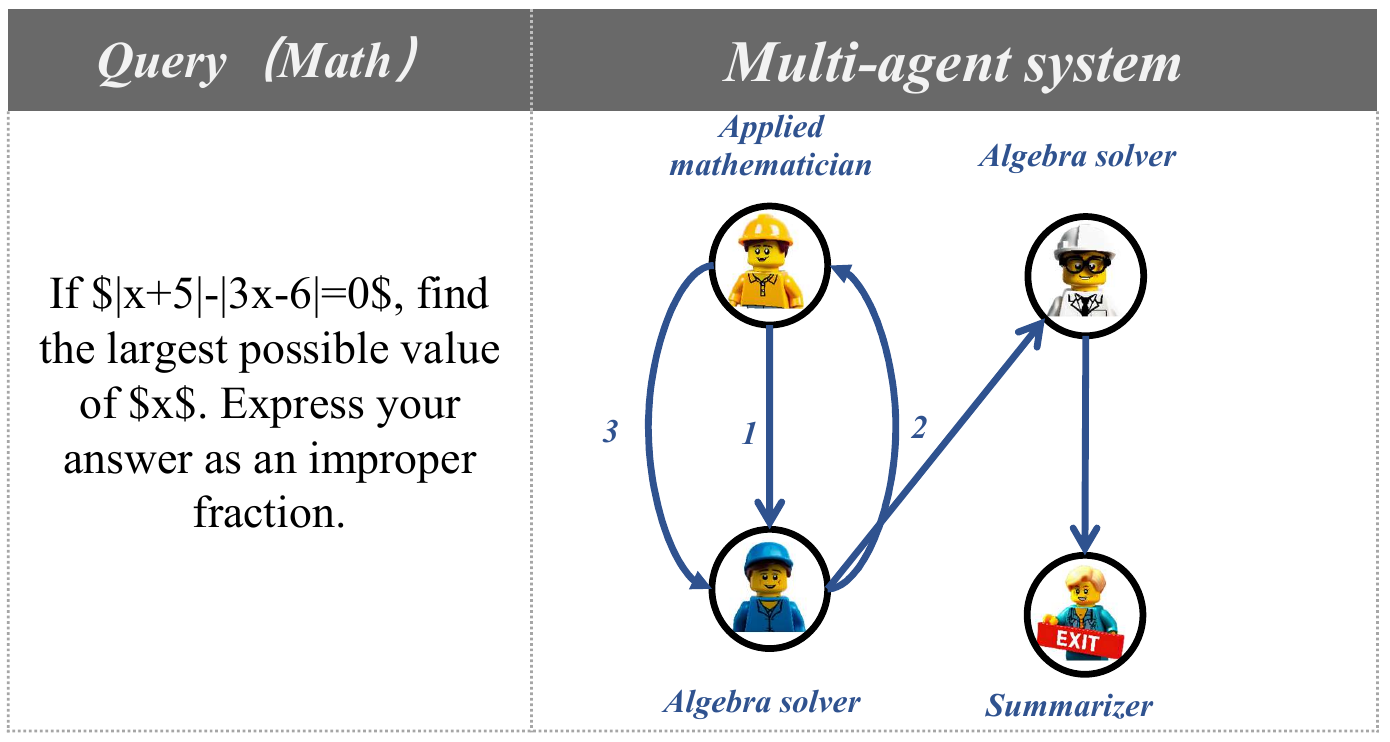}
	\caption{The Mas constructed on the MATH sample.} 
	\label{fig:demo_MATH}
\end{figure}

\begin{figure}
	\centering
	\includegraphics[width=\linewidth]{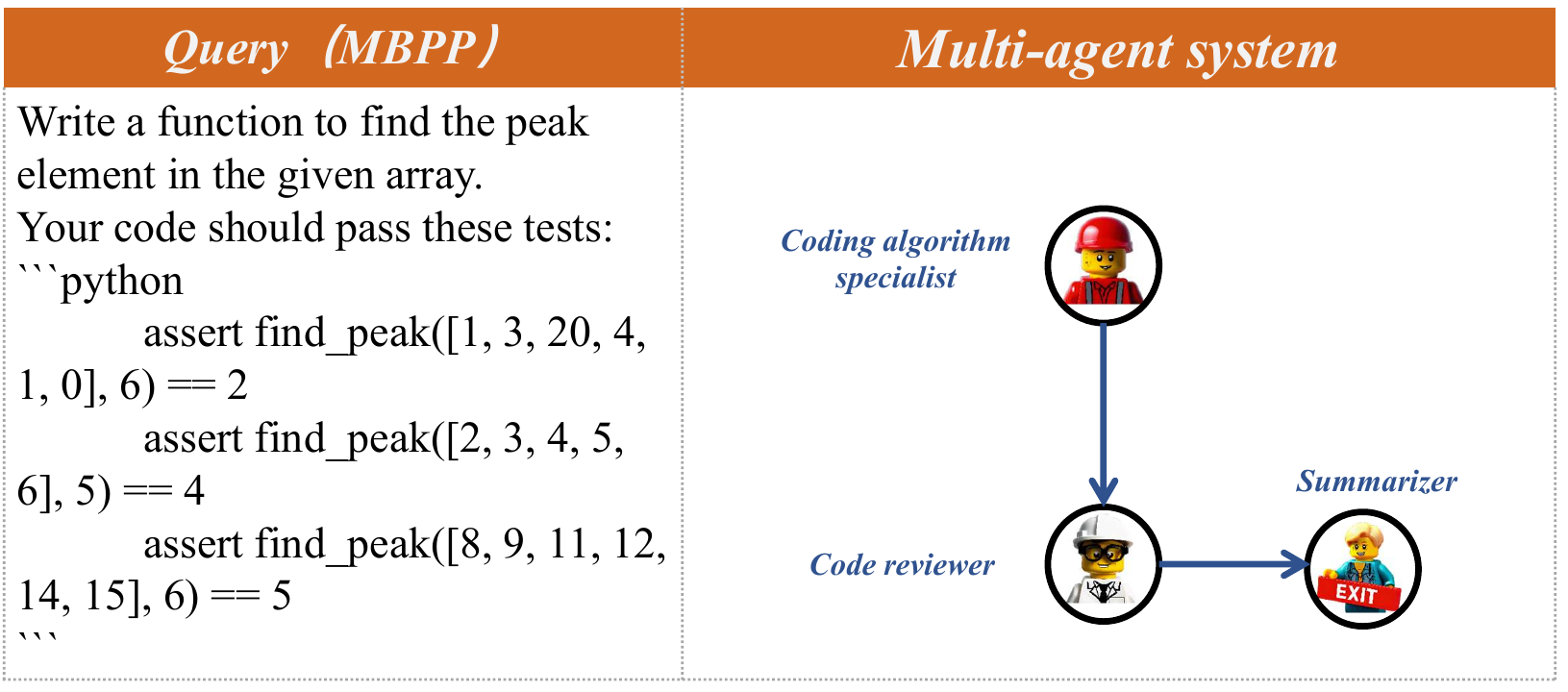}
	\caption{The Mas constructed on the MBPP sample.} 
	\label{fig:demo_MBPP}
\end{figure}

To intuitively demonstrate the effectiveness of our \texttt{MasHost}, we visualize the constructed multi-agent system. As shown in Fig. \ref{fig:demo_GSM8K} \ref{fig:demo_HumanEval} \ref{fig:demo_MATH} \ref{fig:demo_MBPP}, our \texttt{MasHost} yields agents with clearly distinguishable roles and behaviors, offering strong interpretability in both structure and decision-making. The visualized trajectories and interactions not only align well with real-world patterns but also reflect the model's superior performance in terms of coordination and task success. These visualizations compellingly demonstrate that our approach achieves a strong balance between interpretability and performance.

\subsection{Hyper-parameters Settings}
\label{hyper}

The hyper-parameters $\alpha$, $\beta$, $\gamma$, and $\varepsilon$ play a critical balancing role in our framework, mediating trade-offs between reward shaping and learning objectives to ensure stable and effective policy optimization. In this section, we elaborate on their functionality and the specific settings adopted in our implementation.

\begin{itemize}[left=0pt]
	\item The $\alpha$ in Eq. \ref{eq:r_a_T}  is set $0.1$ in the implementation. The $\alpha$ is a balancing hyper-parameter to ensure $- \alpha  \cdot (t-\mathcal{T}_{E}) \in [-1, 0]$. Since the number of exploration steps typically does not exceed 10, the value  $\alpha$ is empirically set to $0.1$. 
	\item The hyperparameter $\beta$ in Eq.~\ref{eq:r_G} is set to $0.0001$ for GSM8K and $0.00001$ for the other datasets. The $\beta$ is a balancing hyper-parameter to ensure $\beta  \cdot {{Tokens}} \in [0, 1]$. The difference setting mainly stems from that the number of tokens consumed per answer in GSM8K ranges from $100$ to $1,000$, whereas in the other datasets, it typically ranges from $1,000$ to $10,000$.
	\item The parameter $\gamma$ in Eq.~\ref{eq:A_hat} is set to $0.9$ in our implementation, following common configurations adopted in reinforcement learning practices. The discount factor $\gamma$ controls the temporal weighting of future rewards, enabling the agent to balance short-term gains with long-term objectives. 
	\item The parameter $\varepsilon$ in Eq.~\ref{eq:loss} is set to $0.1$ in our implementation, following the configuration used in the paper \cite{shao2024deepseekmath, schulman2017proximal}. The clipping threshold $\varepsilon$ constrains policy updates by limiting the change in the probability ratio, thus preventing overly aggressive updates that could destabilize training\cite{schulman2017proximal}.
\end{itemize}

\subsection{Role Prompts}
\begin{figure}[ht]
	\centering
	\begin{subfigure}[b]{0.305\textwidth}
		\centering
		\includegraphics[width=\textwidth]{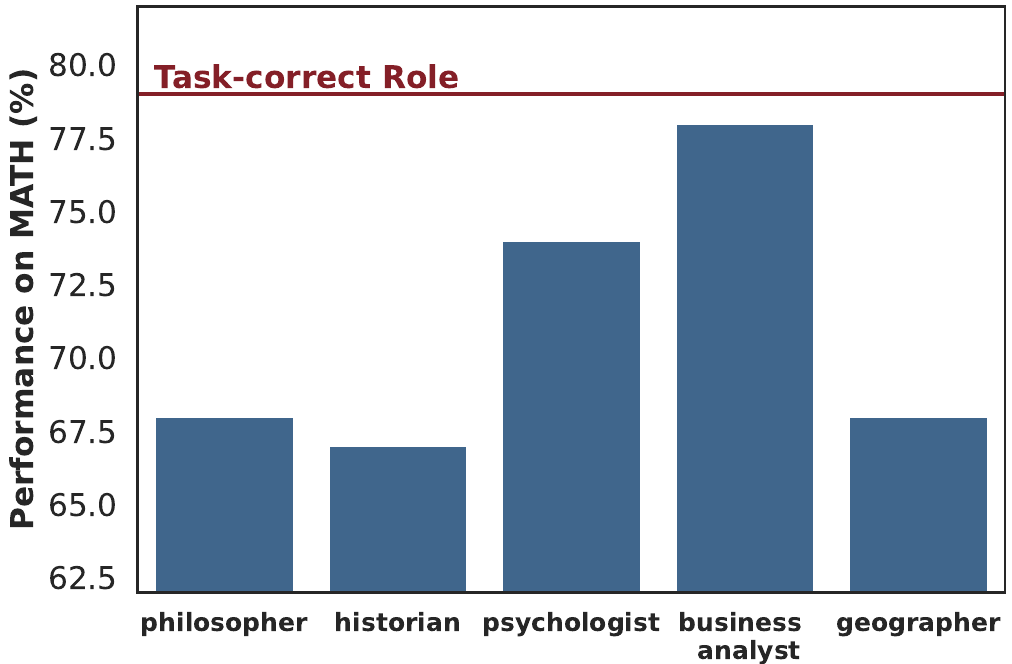}
		\caption{Performance of unreasonable role assignment on MATH.}
	\end{subfigure}
	\begin{subfigure}[b]{0.305\textwidth}
		\centering
		\includegraphics[width=\textwidth]{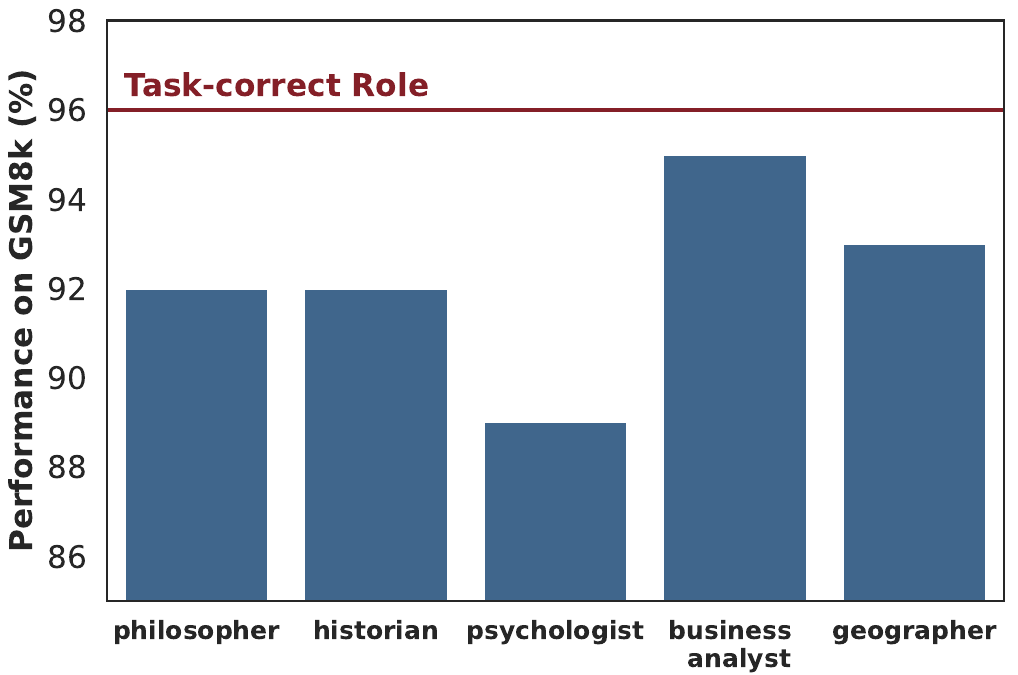}
		\caption{Performance of unreasonable role assignment on GSM8k.}
	\end{subfigure}
	\begin{subfigure}[b]{0.305\textwidth}
		\centering
		\includegraphics[width=\textwidth]{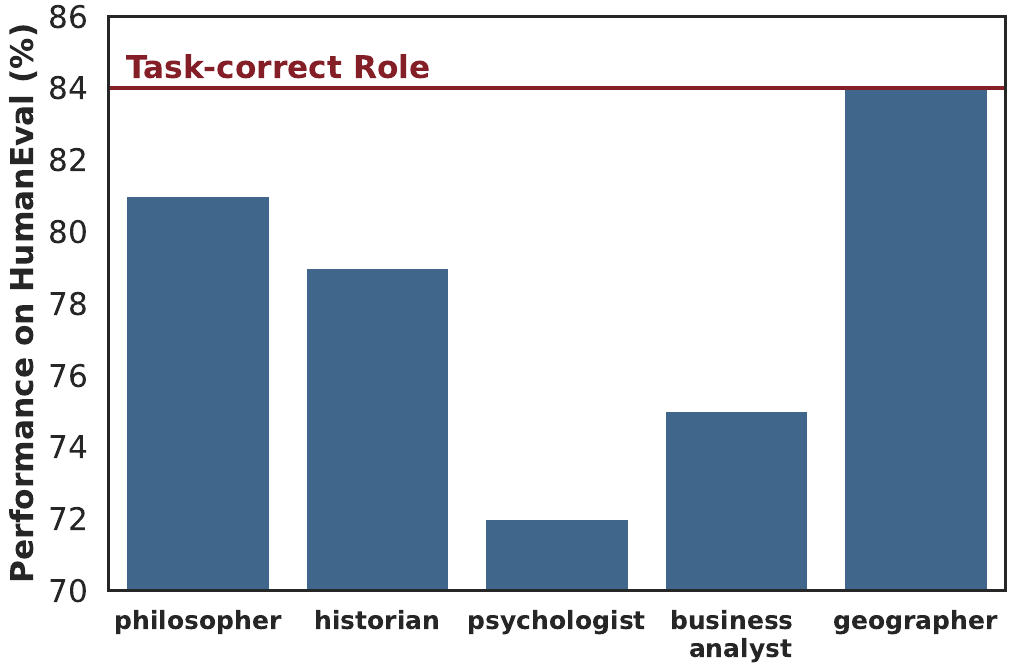}
		\caption{Performance of unreasonable role assignment on HumanEval.}
	\end{subfigure}
	\caption{Roles associated with unrelated tasks are nevertheless able to answer the queries well.}
	\label{fig:Role_Details}
\end{figure}

Our \texttt{MasHost} relies on a global role pool, which includes all known applicable roles. We provide specific role names along with corresponding prompts. Different from existing practices, they overlook the design of refuse conditions for agent. We highlight the specific function and identity of each role agent. This design is motivated by the aim of this work to enhance the rationality of Mas. The irrationality of previous methods lies in their tendency to allow the model to select a role completely unrelated to the question, yet still generate a valid response, as shown in Fig. \ref{fig:Role_Details}. While this may seem acceptable for relatively simple problems, it hinders broader transfer and real-world applicability.

\section{Conclusion}
\label{Conclusion}
In this work, we propose \texttt{MasHost}, a novel reinforcement learning-based framework that enables the fully autonomous construction of query-specific Multi-agent system (Mas).
By introducing a joint probabilistic sampling mechanism and a novel Hierarchical Relative Policy Optimization strategy, \texttt{MasHost} enables end-to-end autonomous design of multi-agent systems with enhanced adaptability, rationality, and performance. Our approach enables scalable, efficient, and interpretable construction of autonomous Mas.

\newpage

\begin{figure}
	\centering
	\includegraphics[width=\linewidth]{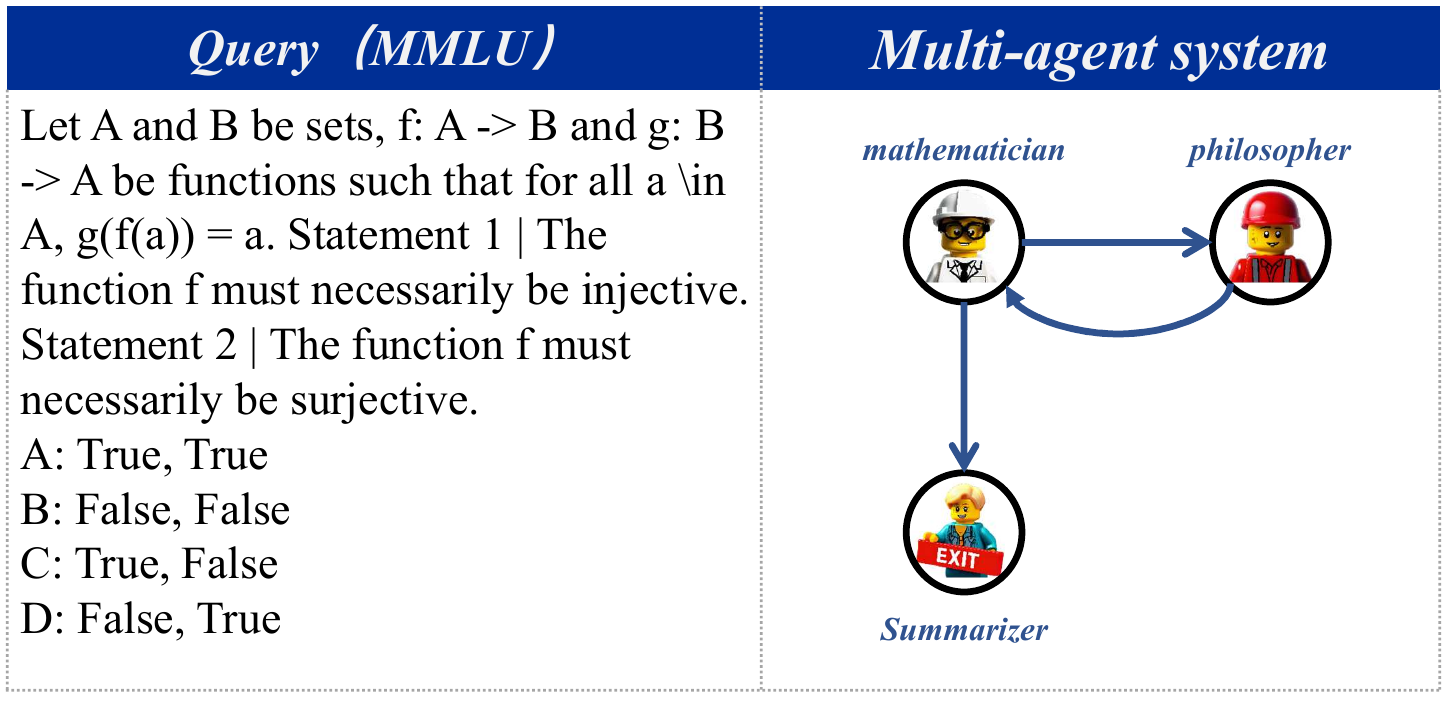}
	\caption{The Mas constructed on the MMLU sample.} 
	\label{fig:demo_MMLU}
\end{figure}

\begin{tcolorbox}[
	colback=gray!10,   
	colframe=gray!80,  
	arc=1mm,           
	boxrule=1pt,
	width=\textwidth,
	sharp corners=downhill, 
	]
	\begin{center}	
		\textcolor{blue!70!black}{\textbf{Molecular Biologist}}\\
	\end{center}
	\textbf{Responsibilities:}
	\begin{itemize}[leftmargin=1.5em]
		\item Study structure and function of biomolecules (DNA, proteins, etc.)
		\item Analyze gene expression and regulation
		\item Investigate molecular mechanisms of cellular processes
		\item Develop techniques like PCR or CRISPR
	\end{itemize}
	\textbf{Assist Conditions:}
	\begin{itemize}[leftmargin=1.5em]
		\item General biology questions
		\item Related fields (e.g., Genetics, Biochemistry, Biotechnology)
	\end{itemize}
\end{tcolorbox}

\begin{tcolorbox}[
	colback=gray!10,   
	colframe=gray!80,  
	arc=1mm,           
	boxrule=1pt,
	width=\textwidth,
	sharp corners=downhill, 
	]
	\begin{center}	
		\textcolor{blue!70!black}{\textbf{Cell Biologist}}\\
	\end{center}
	\textbf{Responsibilities:}
	\begin{itemize}[leftmargin=1.5em]
		\item Study cell structure, division, and metabolism
		\item Investigate cell signaling and communication
		\item Analyze organelle functions (e.g., mitochondria, nucleus)
		\item Research cell responses to environmental changes
	\end{itemize}
	\textbf{Assist Conditions:}
	\begin{itemize}[leftmargin=1.5em]
		\item General biology questions
		\item Related fields (e.g., Molecular Biology, Immunology, Cancer Research)
	\end{itemize}
\end{tcolorbox}

\begin{tcolorbox}[
	colback=gray!10,   
	colframe=gray!80,  
	arc=1mm,           
	boxrule=1pt,
	width=\textwidth,
	sharp corners=downhill, 
	]
	
	\begin{center}	
		\textcolor{blue!70!black}{\textbf{Geneticist}}\\
	\end{center}
	\textbf{Responsibilities:}
	\begin{itemize}[leftmargin=1.5em]
		\item Study inheritance patterns and genetic variation
		\item Analyze DNA sequencing data
		\item Investigate genetic disorders
		\item Develop genetic engineering tools
	\end{itemize}
	\textbf{Assist Conditions:}
	\begin{itemize}[leftmargin=1.5em]
		\item General biology questions
		\item Related fields (e.g., Genomics, Evolutionary Biology, Medicine)
	\end{itemize}

\end{tcolorbox}

\begin{tcolorbox}[
	colback=gray!10,   
	colframe=gray!80,  
	arc=1mm,           
	boxrule=1pt,
	width=\textwidth,
	sharp corners=downhill, 
	]
	
	\begin{center}	
		\textcolor{blue!70!black}{\textbf{Botanist}}\\
	\end{center}
	\textbf{Responsibilities:}
	\begin{itemize}[leftmargin=1.5em]
		\item Study plant physiology and taxonomy
		\item Investigate plant-environment interactions
		\item Research photosynthesis and plant hormones
		\item Explore plant biodiversity and conservation
	\end{itemize}
	\textbf{Assist Conditions:}
	\begin{itemize}[leftmargin=1.5em]
		\item General biology questions
		\item Related fields (e.g., Ecology, Agriculture, Forestry)
	\end{itemize}

\end{tcolorbox}

\begin{tcolorbox}[
	colback=gray!10,   
	colframe=gray!80,  
	arc=1mm,           
	boxrule=1pt,
	width=\textwidth,
	sharp corners=downhill, 
	]
	
	\begin{center}	
		\textcolor{blue!70!black}{\textbf{Biomedical Scientist}}\\
	\end{center}
	\textbf{Responsibilities:}
	\begin{itemize}[leftmargin=1.5em]
		\item Research disease mechanisms (e.g., cancer, infections)
		\item Develop diagnostic tools and therapies
		\item Study drug interactions and pharmacokinetics
		\item Investigate immune system responses
	\end{itemize}
	\textbf{Assist Conditions:}
	\begin{itemize}[leftmargin=1.5em]
		\item General biology questions
		\item Related fields (e.g., Pharmacology, Immunology, Clinical Research)
	\end{itemize}

\end{tcolorbox}

\begin{tcolorbox}[
	colback=gray!10,   
	colframe=gray!80,  
	arc=1mm,           
	boxrule=1pt,
	width=\textwidth,
	sharp corners=downhill, 
	]
	
	\begin{center}	
		\textcolor{blue!70!black}{\textbf{Inorganic Chemist}}\\
	\end{center}
	\textbf{Responsibilities:}
	\begin{itemize}[leftmargin=1.5em]
		\item Study the structure and properties of inorganic compounds
		\item Investigate catalysis and reaction mechanisms in inorganic systems
		\item Develop new materials
		\item Analyze metal-ligand interactions in coordination chemistry
		\item Explore bioinorganic chemistry
	\end{itemize}
	\textbf{Assist Conditions:}
	\begin{itemize}[leftmargin=1.5em]
		\item General chemistry questions
		\item Related fields (e.g., Materials Science, Geochemistry, Industrial Catalysis)
	\end{itemize}

\end{tcolorbox}

\begin{tcolorbox}[
	colback=gray!10,   
	colframe=gray!80,  
	arc=1mm,           
	boxrule=1pt,
	width=\textwidth,
	sharp corners=downhill, 
	]
	
	\begin{center}	
		\textcolor{blue!70!black}{\textbf{Organic Chemist}}\\
	\end{center}
	\textbf{Responsibilities:}
	\begin{itemize}[leftmargin=1.5em]
		\item Study the synthesis, structure, and reactivity of organic compounds
		\item Develop new synthetic methodologies
		\item Investigate reaction mechanisms
		\item Design pharmaceuticals, agrochemicals, or polymers
		\item Analyze spectroscopic data (NMR, IR, MS) for structure elucidation
	\end{itemize}
	\textbf{Assist Conditions:}
	\begin{itemize}[leftmargin=1.5em]
		\item General chemistry questions
		\item Related fields (e.g., Medicinal Chemistry, Polymer Science, Petrochemistry)
	\end{itemize}	
\end{tcolorbox}

\begin{tcolorbox}[
	colback=gray!10,   
	colframe=gray!80,  
	arc=1mm,           
	boxrule=1pt,
	width=\textwidth,
	sharp corners=downhill, 
	]
	
	\begin{center}	
		\textcolor{blue!70!black}{\textbf{Analytical Chemist}}\\
	\end{center}
	\textbf{Responsibilities:}
	\begin{itemize}[leftmargin=1.5em]
		\item Develop and optimize analytical techniques
		\item Perform qualitative and quantitative analysis of chemical samples
		\item Validate methods for quality control
		\item Interpret data from instruments
		\item Ensure compliance with regulatory standards
	\end{itemize}
	\textbf{Assist Conditions:}
	\begin{itemize}[leftmargin=1.5em]
		\item General chemistry questions
		\item Related fields (e.g., Forensic Science, Environmental Monitoring, Food Safety)
	\end{itemize}

\end{tcolorbox}

\begin{tcolorbox}[
	colback=gray!10,   
	colframe=gray!80,  
	arc=1mm,           
	boxrule=1pt,
	width=\textwidth,
	sharp corners=downhill, 
	]
	\begin{center}	
		\textcolor{blue!70!black}{\textbf{Materials Chemist}}\\
	\end{center}
	\textbf{Responsibilities:}
	\begin{itemize}[leftmargin=1.5em]
		\item Design and synthesize novel materials
		\item Study structure-property relationships in materials
		\item Develop functional materials for energy storage
		\item Investigate smart materials
	\end{itemize}
	\textbf{Assist Conditions:}
	\begin{itemize}[leftmargin=1.5em]
		\item General chemistry questions
		\item Related fields (e.g., Nanotechnology, Electronics, Energy Science, Biomedical Engineering)
	\end{itemize}

\end{tcolorbox}

\begin{tcolorbox}[
	colback=gray!10,   
	colframe=gray!80,  
	arc=1mm,           
	boxrule=1pt,
	width=\textwidth,
	sharp corners=downhill, 
	]
	\begin{center}
		\textcolor{blue!70!black}{\textbf{Theoretical Chemist}}\\
	\end{center}
	\textbf{Responsibilities:}
	\begin{itemize}[leftmargin=1.5em]
		\item Develop computational models to predict molecular properties and reactions
		\item Apply quantum mechanics (e.g., DFT, ab initio methods) to chemical systems
		\item Simulate molecular dynamics and statistical mechanics
		\item Analyze chemical bonding and electronic structure
		\item Collaborate with experimentalists to interpret data and guide research
	\end{itemize}
	\textbf{Assist Conditions:}
	\begin{itemize}[leftmargin=1.5em]
		\item General chemistry questions
		\item Related fields (e.g., Computational Chemistry, Drug Design, Catalysis, Astrophysics)
	\end{itemize}

\end{tcolorbox}

\begin{tcolorbox}[
	colback=gray!10,   
	colframe=gray!80,  
	arc=1mm,           
	boxrule=1pt,
	width=\textwidth,
	sharp corners=downhill, 
	]
	\begin{center}
		\textcolor{blue!70!black}{\textbf{Code Reviewer}}\\
	\end{center}
	\textbf{Responsibilities:}
	\begin{itemize}[leftmargin=1.5em]
		\item Analyze code style compliance
		\item Identify potential bugs and security vulnerabilities
		\item Suggest performance optimizations
		\item Evaluate code readability and maintainability
		\item Check boundary conditions and exception handling
	\end{itemize}
	\textbf{Reject Conditions:}
	\begin{itemize}[leftmargin=1.5em]
		\item user mentioned that currently no cooperators available.
		\item Or user gives cooperators, but their messages are not related to code.
	\end{itemize}

\end{tcolorbox}

\begin{tcolorbox}[
	colback=gray!10,   
	colframe=gray!80,  
	arc=1mm,           
	boxrule=1pt,
	width=\textwidth,
	sharp corners=downhill, 
	]
	\begin{center}
		\textcolor{blue!70!black}{\textbf{Code Reviewer}}\\
	\end{center}
	\textbf{Responsibilities:}
	\begin{itemize}[leftmargin=1.5em]
		\item Analyze code style compliance
		\item Identify potential bugs and security vulnerabilities
		\item Suggest performance optimizations
		\item Evaluate code readability and maintainability
		\item Check boundary conditions and exception handling
	\end{itemize}
	\textbf{Reject Conditions:}
	\begin{itemize}[leftmargin=1.5em]
		\item user mentioned that currently no cooperators available.
		\item Or user gives cooperators, but their messages are not related to code.
	\end{itemize}
	
\end{tcolorbox}

\begin{tcolorbox}[
	colback=gray!10,
	colframe=gray!80,
	arc=1mm,
	boxrule=1pt,
	width=\textwidth,
	sharp corners=downhill,
	]
	\begin{center}
		\textcolor{blue!70!black}{\textbf{Debug Assistant}}\\
	\end{center}
	\textbf{Responsibilities:}
	\begin{itemize}[leftmargin=1.5em]
		\item Parse error messages and stack traces
		\item Locate root causes in code
		\item Suggest debugging methods and tools
		\item Verify effectiveness of fixes
		\item Reproduce and isolate error scenarios
	\end{itemize}
	\textbf{Reject Conditions:}
	\begin{itemize}[leftmargin=1.5em]
		\item user mentioned that currently no cooperators available.
		\item Or user gives cooperators, but their messages are not related to code.
	\end{itemize}
	
\end{tcolorbox}

\begin{tcolorbox}[
	colback=gray!10,
	colframe=gray!80,
	arc=1mm,
	boxrule=1pt,
	width=\textwidth,
	sharp corners=downhill,
	]
	\begin{center}
		\textcolor{blue!70!black}{\textbf{Python Programmer}}\\
	\end{center}
	\textbf{Responsibilities:}
	\begin{itemize}[leftmargin=1.5em]
		\item Answer Python language feature questions
		\item Explain standard library and third-party package usage
		\item Guide Python best practices
		\item Analyze advanced features
		\item Compare differences between Python implementations
	\end{itemize}
	\textbf{Assist Conditions:}
	\begin{itemize}[leftmargin=1.5em]
		\item General mathematics or physics questions
	\end{itemize}
	
\end{tcolorbox}

\begin{tcolorbox}[
	colback=gray!10,
	colframe=gray!80,
	arc=1mm,
	boxrule=1pt,
	width=\textwidth,
	sharp corners=downhill,
	]
	\begin{center}
		\textcolor{blue!70!black}{\textbf{Coding Algorithm Specialist}}\\
	\end{center}
	\textbf{Responsibilities:}
	\begin{itemize}[leftmargin=1.5em]
		\item Design optimal algorithms for problems
		\item Analyze time and space complexity
		\item Suggest suitable data structures
		\item Compare different algorithmic approaches
		\item Explain algorithm design patterns
	\end{itemize}
	\textbf{Assist Conditions:}
	\begin{itemize}[leftmargin=1.5em]
		\item General mathematics or physics questions
	\end{itemize}
	
\end{tcolorbox}

\begin{tcolorbox}[
	colback=gray!10,
	colframe=gray!80,
	arc=1mm,
	boxrule=1pt,
	width=\textwidth,
	sharp corners=downhill,
	]
	\begin{center}
		\textcolor{blue!70!black}{\textbf{Performance Optimizer}}\\
	\end{center}
	\textbf{Responsibilities:}
	\begin{itemize}[leftmargin=1.5em]
		\item Identify performance bottlenecks
		\item Suggest low-level optimizations
		\item Analyze memory usage patterns
		\item Guide parallelization strategies
		\item Recommend profiling tools and techniques
	\end{itemize}
	\textbf{Reject Conditions:}
	\begin{itemize}[leftmargin=1.5em]
		\item user mentioned that currently no cooperators available.
		\item Or user gives cooperators, but their messages are not related to code.
	\end{itemize}
	
\end{tcolorbox}

\begin{tcolorbox}[
	colback=gray!10,
	colframe=gray!80,
	arc=1mm,
	boxrule=1pt,
	width=\textwidth,
	sharp corners=downhill,
	]
	\begin{center}
		\textcolor{blue!70!black}{\textbf{Algebra Solver}}\\
	\end{center}
	\textbf{Responsibilities:}
	\begin{itemize}[leftmargin=1.5em]
		\item Solve linear and nonlinear equations
		\item Perform matrix operations and linear algebra computations
		\item Factor and manipulate polynomial expressions
		\item Solve systems of equations
		\item Simplify algebraic expressions
	\end{itemize}
	\textbf{Assist Conditions:}
	\begin{itemize}[leftmargin=1.5em]
		\item General mathematics questions
		\item Related fields (e.g., number theory, geometry)
	\end{itemize}
	
\end{tcolorbox}

\begin{tcolorbox}[
	colback=gray!10,
	colframe=gray!80,
	arc=1mm,
	boxrule=1pt,
	width=\textwidth,
	sharp corners=downhill,
	]
	\begin{center}
		\textcolor{blue!70!black}{\textbf{Geometry Specialist}}\\
	\end{center}
	\textbf{Responsibilities:}
	\begin{itemize}[leftmargin=1.5em]
		\item Explain coordinate geometry concepts
		\item Analyze geometric transformations
		\item Compute areas, volumes and angles
		\item Guide vector geometry applications
		\item Solve trigonometric problems
	\end{itemize}
	\textbf{Assist Conditions:}
	\begin{itemize}[leftmargin=1.5em]
		\item General mathematics questions
		\item Related fields (e.g., Physics applications, Computer graphics, Architectural design)
	\end{itemize}
	
\end{tcolorbox}

\begin{tcolorbox}[
	colback=gray!10,
	colframe=gray!80,
	arc=1mm,
	boxrule=1pt,
	width=\textwidth,
	sharp corners=downhill,
	]
	
	\begin{center}
		\textcolor{blue!70!black}{\textbf{Applied Mathematician}}\\
	\end{center}
	\textbf{Responsibilities:}
	\begin{itemize}[leftmargin=1.5em]
		\item Bridge theoretical math and practical applications
		\item Solve mathematical modeling problems
		\item Explain numerical analysis methods
		\item Guide optimization problem solutions
		\item Analyze operations research problems
	\end{itemize}
	\textbf{Assist Conditions:}
	\begin{itemize}[leftmargin=1.5em]
		\item General mathematics questions
		\item Related fields (e.g., Engineering problems, Economic modeling, Scientific computing)
	\end{itemize}
	
\end{tcolorbox}

\begin{tcolorbox}[
	colback=gray!10,
	colframe=gray!80,
	arc=1mm,
	boxrule=1pt,
	width=\textwidth,
	sharp corners=downhill,
	]
	\begin{center}
		\textcolor{blue!70!black}{\textbf{Analytic Mathematician}}\\
	\end{center}
	\textbf{Responsibilities:}
	\begin{itemize}[leftmargin=1.5em]
		\item Study limits, continuity, and convergence in real and complex spaces
		\item Develop theories in calculus, measure theory, and functional analysis
		\item Solve differential equations and harmonic analysis problems
		\item Explore Fourier analysis and operator theory
		\item Investigate partial differential equations and their applications
	\end{itemize}
	\textbf{Assist Conditions:}
	\begin{itemize}[leftmargin=1.5em]
		\item General mathematics questions
		\item Related fields (e.g., Mathematical physics, Dynamical systems, Probability theory)
	\end{itemize}
	
\end{tcolorbox}

\begin{tcolorbox}[
	colback=gray!10,
	colframe=gray!80,
	arc=1mm,
	boxrule=1pt,
	width=\textwidth,
	sharp corners=downhill,
	]
	\begin{center}
		\textcolor{blue!70!black}{\textbf{Discrete Mathematician}}\\
	\end{center}
	\textbf{Responsibilities:}
	\begin{itemize}[leftmargin=1.5em]
		\item Study combinatorial structures and graph theory
		\item Solve problems in cryptography and coding theory
		\item Analyze discrete optimization and algorithmic complexity
		\item Explore logic, set theory, and discrete probability
		\item Investigate network science and computational geometry
	\end{itemize}
	\textbf{Assist Conditions:}
	\begin{itemize}[leftmargin=1.5em]
		\item General mathematics questions
		\item Related fields (e.g., Computer science, Cryptography, Operations research)
	\end{itemize}
	
\end{tcolorbox}

\begin{tcolorbox}[
	colback=gray!10,
	colframe=gray!80,
	arc=1mm,
	boxrule=1pt,
	width=\textwidth,
	sharp corners=downhill,
	]
	\begin{center}
		\textcolor{blue!70!black}{\textbf{Classical Physicist}}\\
	\end{center}
	\textbf{Responsibilities:}
	\begin{itemize}[leftmargin=1.5em]
		\item Study macroscopic physics (mechanics, thermodynamics, electromagnetism)
		\item Analyze motion and forces in Newtonian frameworks
		\item Model wave phenomena and fluid dynamics
		\item Explain classical field theories
	\end{itemize}
	\textbf{Assist Conditions:}
	\begin{itemize}[leftmargin=1.5em]
		\item General physics questions
		\item Related fields (e.g., Engineering mechanics, Acoustics, Thermodynamic systems)
	\end{itemize}
	
\end{tcolorbox}

\begin{tcolorbox}[
	colback=gray!10,
	colframe=gray!80,
	arc=1mm,
	boxrule=1pt,
	width=\textwidth,
	sharp corners=downhill,
	]
	\begin{center}
		\textcolor{blue!70!black}{\textbf{Particle Physicist}}\\
	\end{center}
	\textbf{Responsibilities:}
	\begin{itemize}[leftmargin=1.5em]
		\item Investigate fundamental particles and interactions
		\item Interpret data from colliders (e.g., LHC)
		\item Test predictions of the Standard Model
		\item Explore beyond-Standard-Model theories
	\end{itemize}
	\textbf{Assist Conditions:}
	\begin{itemize}[leftmargin=1.5em]
		\item General physics questions
		\item Related fields (e.g., Quantum field theory, Cosmology, Nuclear physics)
	\end{itemize}
	
\end{tcolorbox}

\begin{tcolorbox}[
	colback=gray!10,
	colframe=gray!80,
	arc=1mm,
	boxrule=1pt,
	width=\textwidth,
	sharp corners=downhill,
	]
	\begin{center}
		\textcolor{blue!70!black}{\textbf{Quantum Physicist}}\\
	\end{center}
	\textbf{Responsibilities:}
	\begin{itemize}[leftmargin=1.5em]
		\item Study quantum systems and entanglement
		\item Develop quantum computing/algorithms
		\item Analyze atomic/subatomic behavior
		\item Explain quantum measurement problems
	\end{itemize}
	\textbf{Assist Conditions:}
	\begin{itemize}[leftmargin=1.5em]
		\item General physics questions
		\item Related fields (e.g., Quantum chemistry, Nanotechnology, Quantum optics)
	\end{itemize}
	
\end{tcolorbox}

\begin{tcolorbox}[
	colback=gray!10,
	colframe=gray!80,
	arc=1mm,
	boxrule=1pt,
	width=\textwidth,
	sharp corners=downhill,
	]
	\begin{center}
		\textcolor{blue!70!black}{\textbf{Condensed Matter Physicist}}\\
	\end{center}
	\textbf{Responsibilities:}
	\begin{itemize}[leftmargin=1.5em]
		\item Research solid/liquid state properties
		\item Study superconductivity or topological materials
		\item Model phase transitions and collective phenomena
		\item Design novel materials (e.g., graphene)
	\end{itemize}
	\textbf{Assist Conditions:}
	\begin{itemize}[leftmargin=1.5em]
		\item General physics questions
		\item Related fields (e.g., Semiconductor physics, Materials science, Spintronics)
	\end{itemize}
	
\end{tcolorbox}

\begin{tcolorbox}[
	colback=gray!10,
	colframe=gray!80,
	arc=1mm,
	boxrule=1pt,
	width=\textwidth,
	sharp corners=downhill,
	]
	\begin{center}
		\textcolor{blue!70!black}{\textbf{Relativistic Physicist}}\\
	\end{center}
	\textbf{Responsibilities:}
	\begin{itemize}[leftmargin=1.5em]
		\item Analyze spacetime curvature (GR effects)
		\item Model black hole/neutron star dynamics
		\item Test Lorentz invariance and relativistic jets
		\item Simulate gravitational wave sources
	\end{itemize}
	\textbf{Assist Conditions:}
	\begin{itemize}[leftmargin=1.5em]
		\item General physics questions
		\item Related fields (e.g., Astrophysics, Cosmology, High-energy physics)
	\end{itemize}
	
\end{tcolorbox}


\newpage
\bibliographystyle{plain}
\bibliography{mashost}


\end{document}